\documentclass[conference]{IEEEtran}
\IEEEoverridecommandlockouts
\usepackage{cite}
\usepackage{amsmath,amssymb,amsfonts}
\usepackage{algorithmic}
\usepackage{textcomp}
\usepackage{bbding}
\usepackage{makecell}
\usepackage{algorithm}  
\usepackage{color}
\usepackage{multirow}
\usepackage{array}
\usepackage{booktabs}
\usepackage{tabularx,booktabs}
\usepackage{multicol}
\usepackage{cleveref}
\usepackage{balance}
\usepackage{lastpage}
\usepackage{tabulary}
\usepackage{subfigure}
\usepackage{etoolbox}
\usepackage{bbm}
\usepackage{mathrsfs}
\usepackage{tablefootnote}
\usepackage{threeparttable}
\usepackage{graphicx} 
\usepackage{epstopdf}

\usepackage{fancyhdr}

\usepackage{mathtools}
\usepackage[x11names]{xcolor}
\newtagform{blue}{\color{black}(}{)}

\def\BibTeX{{\rm B\kern-.05em{\sc i\kern-.025em b}\kern-.08em
    T\kern-.1667em\lower.7ex\hbox{E}\kern-.125emX}}

\begin{document}

\title{\vspace{-13pt} Learning from Peers: Deep Transfer Reinforcement Learning for Joint Radio and Cache Resource Allocation in 5G RAN Slicing\\}

\author{\IEEEauthorblockN{Hao Zhou, \IEEEmembership{Graduate Student Member, IEEE}, Melike Erol-Kantarci, \IEEEmembership{Senior Member, IEEE}, Vincent Poor, \IEEEmembership{Fellow, IEEE} }

\thanks{ This work was supported by Natural Sciences and Engineering Research Council of Canada (NSERC) Collaborative Research and Training Experience Program (CREATE) under Grant 497981, Canada Research Chairs Program, and U.S. National Science Foundation under Grant CNS-2128448. 

H. Zhou and M. Erol-Kantarci are with the School of Electrical Engineering and Computer Science, University of Ottawa, Ottawa, ON K1N 6N5, Canada. (emails:\{hzhou098, melike.erolkantarci\}@uottawa.ca).

H. V. Poor is with the Department of Electrical and Computer Engineering,
Princeton University, Princeton, NJ 08544 USA (e-mail: poor@princeton.edu).}}

\maketitle
\thispagestyle{fancy}            
\chead{This paper has been accepted by IEEE Transactions on Cognitive Communications and Networking }                     
\pagestyle{plain}                

\begin{abstract}
Network slicing is a critical technique for 5G communications that covers radio access network (RAN), edge, transport and core slicing. 
The evolving network architecture requires the orchestration of multiple network resources such as radio and cache resources. In recent years, machine learning (ML) techniques have been widely applied for network management. However, most existing works do not take advantage of the knowledge transfer capability in ML. In this paper, we propose a deep transfer reinforcement learning (DTRL) scheme for joint radio and cache resource allocation to serve 5G RAN slicing. We first define a hierarchical architecture for joint resource allocation. Then we propose two DTRL algorithms: Q-value-based deep transfer reinforcement learning (QDTRL) and action selection-based deep transfer reinforcement learning (ADTRL). In the proposed schemes, learner agents utilize expert agents' knowledge to improve their performance on current tasks. The proposed algorithms are compared with both the model-free exploration bonus deep Q-learning (EB-DQN) and the model-based priority proportional fairness and time-to-live (PPF-TTL) algorithms. Compared with EB-DQN, our proposed DTRL-based method presents 21.4\% lower delay for Ultra Reliable Low Latency Communications (URLLC) slice and 22.4\% higher throughput for enhanced Mobile Broad Band (eMBB) slice, while achieving significantly faster convergence than EB-DQN. Moreover, 40.8\% lower URLLC delay and 59.8\% higher eMBB throughput are observed with respect to PPF-TTL. 
\end{abstract}


\section{Introduction}
Driven by the increasing traffic demand of diverse mobile applications, 5G mobile networks are expected to satisfy the diverse quality of service (QoS) requirements of a wide variety of services as well as service level agreements of different user types \cite{b1}. Considering the diverse QoS demands of user types such as enhanced Mobile Broad Band (eMBB) and Ultra Reliable Low Latency Communications (URLLC), network slicing has been proposed to enable flexibility and customization of 5G networks. Based on software defined networks and network function virtualization techniques, physical networks are split into multiple logical network slices\cite{b2}. Each slice may include its own controller to manage the available resources.          

As an important part of network slicing, radio access network (RAN) slicing is more complicated than core and transport network slicing due to limited bandwidth resources and fluctuating radio channels. For instance, a two-layer slicing approach is introduced in \cite{b4} for low complexity RAN slicing, which aims to find a suitable trade-off between slice isolation and efficiency. A puncturing-based scheduling method is proposed in \cite{b5} to allocate resources for the incoming URLLC traffic and minimize the risk of interrupting eMBB transmissions. However, the puncturing-based method may degrade the performance of the eMBB slice, since the URLLC slice is scheduled on top of ongoing eMBB transmissions (i.e., puncturing the current eMBB transmission).

The aforementioned works mainly concentrate on RAN slicing. While the spectrum is indisputably the most critical resource of RAN, other resources are equally important to guarantee the network performance, especially cache resource \cite{b6}. Incorporating caching into the RAN has attracted interest from both academia and industry, and some novel network architectures have been proposed to harvest the potential advantages of edge caching\cite{b6-2}. Indeed, edge caching allows storing data closer to the users by utilizing the storage capacity available at the network devices, and it reduces the traffic to the core network. Moreover, caching can save the backhaul capacity without affecting the network delay. Cache placement and caching strategies have been extensively studied in numerous works, but the problem is usually addressed disjointly without considering the RAN. For example, a location customized caching method is presented in \cite{b8} to maximize the cache hit ratio, and content caching locations are optimized in \cite{b9} by considering both cloud-centric and edge-centric caching.

Augmenting edge devices in RAN with caching will bring significant improvements for 5G networks. However, this leads to higher complexity for network management. Network slicing is expected to allocate limited resources such as bandwidth or caching capacity between slices and fulfill the QoS requirements of slices. The complex network dynamics, especially the stochastic arrival requests of slices, make the underlying network optimization challenging. Fortunately, machine learning (ML) techniques offer promising solutions \cite{b10}. Applying a reinforcement learning (RL) scheme can avoid the potential complexity of defining a dedicated optimization model. 
For instance, Q-learning is deployed in \cite{b12} to maximize the network utility of 5G RAN slicing by jointly considering radio and computation resources. Deep Q-learning (DQN) is used in \cite{b13} for end-to-end network slicing, and double deep Q-learning (DDQN) is applied for computation offloading within sliced RANs in \cite{b14}.

Although learning-based methods such as RL and deep reinforcement learning (DRL) have been generally applied for network resource allocation, most existing works do not consider the possibility of knowledge transfer \cite{b12,b13,b14}. Specifically, an agent is designed for a specific task in these works, and it interacts with its environment from scratch, which usually leads to a lower exploration efficiency and longer convergence time. Whenever a new task is assigned, the agent needs to be retrained, even though similar tasks have been completed before. The poor generalization capability of straightforward RL methods motivates us to find a learning method with better generalization and knowledge transfer capability. On the other hand, humans can reuse the knowledge learned from previous tasks to solve new tasks more efficiently, and this capability can be built into ML as well\cite{b23}. Such knowledge transfer and reuse can significantly reduce the need for a large number of training samples, which is a common issue in many ML methods. By incorporating knowledge transfer capability into ML, it is expected to reduce the algorithm design and training efforts and achieve better performance such as faster convergence and higher average reward.

In this work, we propose two deep transfer reinforcement learning (DTRL) based solutions for joint radio and cache resource allocation. In particular, we include knowledge transfer capability in the DDQN framework by defining two different knowledge transfer functions, and we propose two DTRL-based algorithms accordingly. The first method is Q-value-based deep transfer reinforcement learning (QDTRL), and the second technique is called action selection-based deep transfer reinforcement learning (ADTRL). Using these schemes, agents can utilize the knowledge of experts to improve their performance on current tasks, and consequently it can reduce the algorithm training efforts. Furthermore, the current network optimization schemes are usually defined in a centralized way\cite{b12,b13}. This leads to excessive control overhead where processing the specific requests of all devices can be a heavy burden for the central controller. To this end, we propose a hierarchical architecture for joint resource allocation. The global resource manager (GRM) is responsible for inter-slice resource allocation, and then slice resource managers (SRMs) implement intra-slice resource allocation among associated user equipment (UEs).

The main contributions of this work are: (1) We define a hierarchical architecture for joint radio and cache resource allocation of cellular networks. The GRM will intelligently allocate resources between slices, then SRMs will distribute radio and cache resource within each slice based on specific rules. The proposed hierarchical architecture can reduce the control overhead and achieve higher flexibility. 

(2) We propose two DTRL-based solutions for the inter-slice resource allocation, namely QDTRL and ADTRL. The QDTRL can utilize the Q-values of the experts as prior knowledge to improve the learning process, while the ADTRL focuses on action selection knowledge. Compared with RL or DRL, the proposed DTRL solutions show a better knowledge transfer capability.

We further propose two baseline algorithms, including a model-free exploration bonus DQN (EB-DQN) algorithm and a model-based priority proportional fairness and time-to-live (PPF-TTL) method. The proposed DTRL solutions are compared with these two baseline algorithms via simulations.
The results demonstrate that DTRL-based algorithms perform better in both network and ML metrics. In particular, ADTRL has 21.4\% lower URLLC delay and 22.4\% higher eMBB throughput than EB-DQN. The simulations also show 40.8\% lower URLLC delay and 59.8\% higher eMBB throughput than the PPF-TTL method. QDTRL and ADTRL also outperform EB-DQN with significantly faster convergence.  

The rest of this work is organized as follows. Section \ref{s2} presents related work, Section \ref{s2-1} introduces the background, and Section \ref{s3} shows the system model and problem formulation. Section \ref{s4} explains the DTRL-based resource allocation scheme. The simulations are shown in Section \ref{s5}, and Section \ref{s6} concludes this work. 

\begin{table*}[!t]
\caption{Comparison of RL, TL, DRL and TRL}
\centering
\label{tab-1}
\renewcommand\arraystretch{1.4}
\begin{tabular}{|m{1.5cm}<{\centering}|m{6cm}<{\centering}|m{4.5cm}<{\centering}|m{4cm}<{\centering}|}
\hline
Algorithms & Features  & Difficulties & Applications\\
\hline
RL & Agent has no prior knowledge about tasks. It explores new tasks from scratch.  & Long convergence time, low exploration efficiency. &  Tasks with limited state-action space and no prior knowledge.\\
\hline
TL  & Improving generalization across different distributions between expert and learner tasks. &  Negative transfer, automation of task mapping. & It is mainly designed for the supervised learning domain, e.g., classification, regression.\\
\hline
DRL  & Combining artificial neural networks with RL architecture. Applying neural networks to estimate state-action values. & Time-consuming network training and tuning; Low sample and exploration efficiency; Training stability. & Large state-action space or continuous-action problem.\\
\hline
TRL & Utilizing prior knowledge from experts to improve performance of learners such as higher average reward or faster convergence. & Transfer function needs to be defined to digest prior knowledge. & Optimization of tasks with existing prior knowledge. \\
\hline
\end{tabular}
\end{table*}

\section{Related Work} 
\label{s2}
Recently, numerous studies have applied artificial intelligence (AI) techniques for resource allocation in 5G networks. DRL is deployed in \cite{b16} for spectrum allocation in integrated access and backhaul networks with dynamic environments, and DRL is combined with game theory in \cite{b16-2} for multi-tenant cross-slice resource allocation. Multi-agent reinforcement learning is used in \cite{b17} for distributed dynamic spectrum access sharing of communication networks. Moreover, a DRL-based method is proposed in \cite{b18} for mobility-aware proactive resource allocation, which pre-allocates resources to mobile UEs in time and frequency domains. Finally, \cite{b19} proposed a DQN-based intelligent resource management method to improve the quality of service for 5G cloud RAN.
The aforementioned works show that various ML methods have been applied for the resource management of wireless networks, including RL\cite{b12}, DRL\cite{b13,b16,b16-2,b18,b19}, DDQN\cite{b14}, multi-agent reinforcement learning \cite{b17}, and so on. In our former work \cite{b20-0}, we proposed a correlated Q-learning-based method for the radio resource allocation of 5G RAN, but the knowledge transfer was still not considered.

In these works, the main motivations for deploying ML algorithms are the increasing complexity of wireless networks and the difficulties to build dedicated optimization models. Indeed, the evolving network architecture, emerging new network performance requirements and increasing device numbers make the traditional methods such as convex optimization more and more complicated. ML, especially RL, offers a good opportunity to reduce the optimization complexity and make data-driven decisions. However, a large number of samples are needed for the learning process, which means a long training time and low system efficiency. Furthermore, after the long learning process, the agent can only handle one specific task without any generalization. This learning process has to be repeated when a new task is assigned. DRL can be considered as a solution to reduce the number of samples and improve convergence. However, it is still limited to a localized domain, and the neural network training can be time-consuming because of the hyperparameter tuning. To this end, we propose DTRL-based solutions for the joint radio and cache resource allocation of 5G RAN. The transfer reinforcement learning (TRL) has been applied in \cite{b20-1} for intra-beam interference management of 5G mm-Wave, but here, the resource allocation problem is more complicated due to a much bigger state-action space. The proposed scheme has a satisfying knowledge transfer capability by utilizing the knowledge of expert agents, and it outperforms EB-DQN by faster convergence and better network performance.

\section{Background}
\label{s2-1}
\begin{figure}[!h]
\centering
\includegraphics[width=9cm,height=3.5cm]{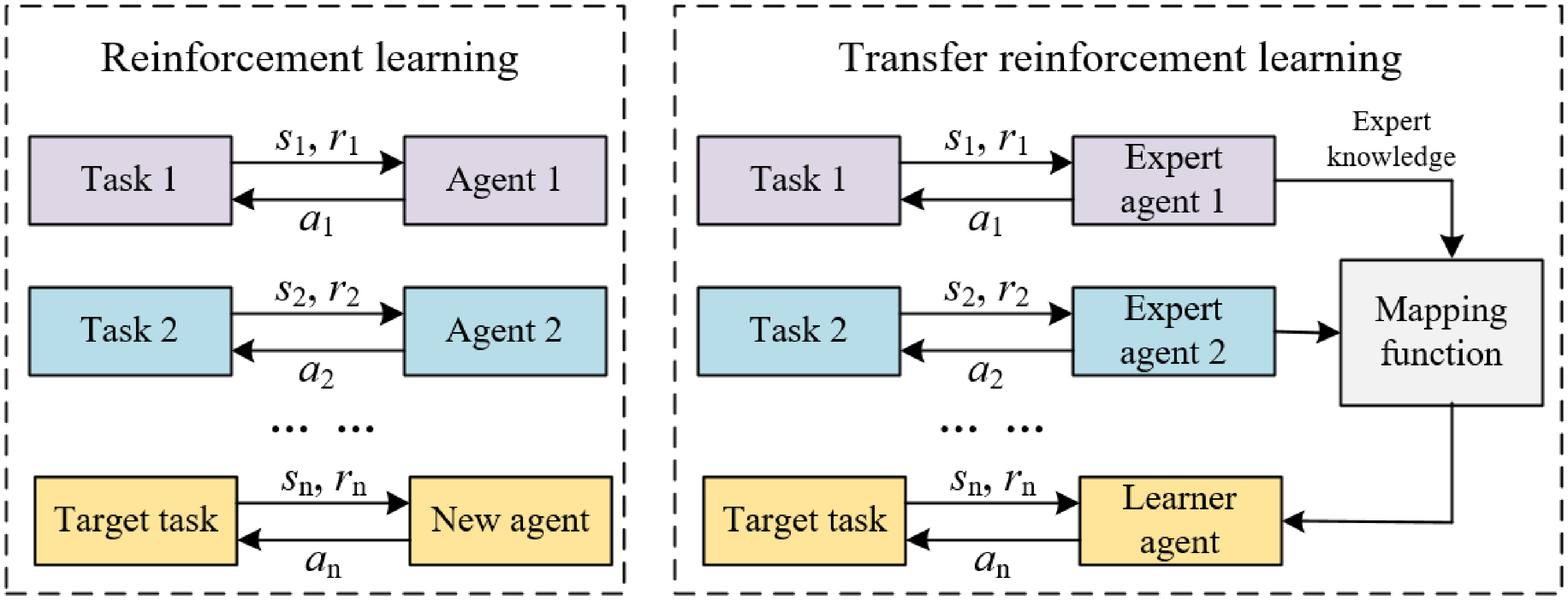}
\caption{Comparison of TRL and RL.}
\label{fig1}
\end{figure}  

In this section, we introduce the background of TRL to distinguish it from RL. As shown in Fig.\ref{fig1}, given a task pool, the interactions between tasks and agents can be described by the Markov decision process (MDP) $<S,A,R,T>$, where $S$ is the state space, $A$ is the action space, $R$ is the reward function, and $T$ is the transition probability. In RL, each agent works independently to try different actions, arriving at new states and receiving rewards. The learning phase of RL can be defined as:
\begin{equation} \label{eq-1}
\mathscr{L}_{RL}:s \times \mathscr{K}\rightarrow a, r \,(a\in A),
\end{equation}
where the $\mathscr{K}$ is the knowledge of this agent, $s$ is the current state, $a$ is the selected action, and $r$ is the reward. Equation (\ref{eq-1}) indicates that RL agent utilizes the collected knowledge to select action $a$ and receive reward $r$ under state $s$.  

On the contrary, TRL includes two phases: the knowledge transfer phase and the learning phase. In the knowledge transfer phase, as shown in Fig.\ref{fig1}, considering task differences, a mapping function is defined to make the knowledge of experts digestible for the learner. Then the learner explores current tasks on its own and forms its own knowledge. The whole process of TRL can be defined as:      
\begin{equation} \label{eq-2}
\mathscr{L}_{TRL}:s \times \mathscr{M}(\mathscr{K}_{expert}) \times \mathscr{K}_{learner} \rightarrow a,r \,(a\in A),
\end{equation}
where the $\mathscr{M}$ is the mapping function, $\mathscr{K}_{expert}$ is the knowledge from experts, $\mathscr{K}_{learner}$ is the knowledge of the learner. In equation (\ref{eq-2}), knowledge from experts is utilized in the action selection of the learner, and it is expected to accelerate the learning process $\mathscr{L}_{TRL}$. In TRL, new tasks can be better handled based on knowledge of experts.

The RL, transfer learning (TL), DRL and TRL techniques are compared in Table \ref{tab-1}. Compared with RL, TRL has higher exploration efficiency and better generalization capability \cite{b21}. On the other hand, although DRL is a breakthrough approach by combining neural networks with RL schemes, the time-consuming network training is a well-known issue, and the training stability and generalization capability of DRL can cause problems \cite{b22}. Finally, although TL has been extensively studied in the ML literature, it mainly focuses on the supervised learning domain such as classification and regression\cite{b23}. Compared with TL, TRL can be more complicated because the knowledge needs to be transferred in the context of the MDP scheme. Moreover, due to the dedicated components of MDP, the knowledge may exist in different forms, which needs to be transferred in different ways\cite{b24}. Furthermore, TRL has shown significant improvements in robot learning. Inspired by these approaches and previous successful results \cite{b20-1}, we propose DTRL-based schemes for RAN slicing. 

In this work, we combine state-of-the-art DDQN with TRL and propose a DTRL-based joint resource allocation method for 5G RAN slicing. Compared with conventional DRL that explores the task from scratch, the DTRL can extract the prior knowledge of related tasks and reuse it for target tasks.

\section{System Model and Problem Formulation}
\label{s3}
\subsection{Overall Architecture}
\begin{figure}[!t]
\centering
\includegraphics[width=8.5cm,height=6.7cm]{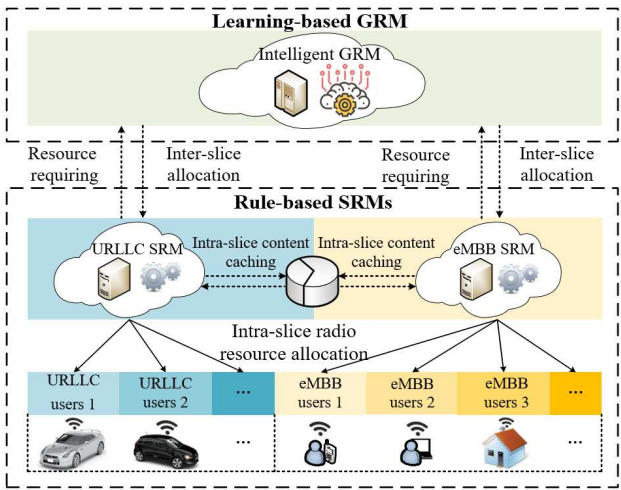}
\caption{Proposed radio and cache resource allocation scheme}
\label{fig1-1}
\vspace{-5pt}
\end{figure}

Fig.\ref{fig1-1} presents the proposed joint radio and cache resource allocation scheme. A hierarchical architecture is defined for the resource allocation of cache-enabled macro cellular networks. We assume the base station (BS) has the caching capability to store content items. Our proposed scheme can be used for any number of slices without loss of generality, but here we mainly consider two typical slices, namely an eMBB slice and URLLC slice, to better illustrate the framework. Firstly, the SRMs collect the QoS requirements of associated UEs, and the collected information is sent to the GRM. Then the GRM will intelligently implement the inter-slice resource allocation to divide the radio and cache resource between eMBB and URLLC SRMs. Finally, SRMs distribute resource blocks (RBs) to associated UEs and update cached content items within the allocated caching capacity. We apply learning-based methods to realize an intelligent GRM, while rule-based methods are deployed for SRMs. This hierarchical architecture can alleviate the burden of GRM, since the GRM only accounts for slice-level allocation instead of handling all the UEs directly.   

\subsection{Communication Model}
In this section, we introduce the communication model. Firstly, the total delay consists of:
\begin{equation} \label{eq1}
d_{j,m,g}=d^{tx}_{j,m,g}+d^{que}_{j,m,g}+(1-\beta_{j,g})d^{back}_{j,m,g},
\end{equation}
where $d^{tx}_{j,m,g}$ is the transmission delay of content item $g$ from BS $j$ to UE $m$. $d^{que}_{j,m,g}$ refers to the queuing delay that content $g$ waits in the buffer of BS $j$ to be transmitted to UE $m$. $d^{back}_{j,m,g}$ is the backhaul delay of fetching content items from the core network. $\beta_{j,g}$ is a binary variable. $\beta_{j,g}=1$ when the content item $g$ is cached at BS $j$, and $\beta_{j,g}=0$ otherwise. Equation (\ref{eq1}) shows that the total delay is affected by both communication and cache resource. The transmission delay depends on radio resource allocation, and the edge caching can prevent the backhaul delay. The scheduling efficiency will affect the queuing delay. 

The transmission delay depends on the link capacity between the BS and UE:
\begin{equation} \label{eq1-1}
d^{tx}=\frac{L_{m}}{P_{j,m}},
\end{equation}
where $L_{m}$ is the size of content items required by UE $m$. $P_{j,m}$ is the link capacity between BS $j$ and UE $m$, which is calculated as follows:
\begin{equation}
\label{eq2}
P_{j,m}=\sum _{q\in{\mathcal{N}^{RB}_{j}}}b_{q}log(1+ \frac{p_{j,q}x_{j,q,m}g_{j,q,m}}{b_{q}N_{0}+\sum\limits_{j'\in J_{-j}}{p_{j',q'}x_{j',q',m'}g_{j',q',m'}}}),
\end{equation}
where $\mathcal{N}^{RB}_{j}$ is the set of RBs in BS $j$, $b_{q}$ is the bandwidth of RB $q$, $N_{0}$ is the noise power density, $p_{j,q}$ is the transmission power of the RB $q$ of the BS $j$, $x_{j,q,m}$ is a binary indicator to denote whether RB $q$ is allocated to the UE $m$, $g_{j,q,m}$ is the channel gain between the BS and UE, and $j'\in J_{-j}$ is the BS set except BS $j$. In the proposed communication system model, we assume orthogonal frequency-division multiplexing (OFDM) is deployed to avoid the intra-cell interference\cite{b25-0}, and $\sum\limits_{j'\in J_{-j}}{p_{j',q'}x_{j',q',m'}g_{j',q',m'}}$ in equation (\ref{eq2}) indicates the inter-cell interference of downlink transmission from other BSs\cite{b25-1}.

\subsection{Slicing-based Caching Model}
We will introduce the slicing-based caching model in this section, in which the time-to-live (TTL) method is used as the content replacement strategy. The TTL indicates the time that a content item is stored in a caching system before it is deleted or replaced. The TTL value will be reset if this content item is required again, and thus popular content items will live longer. Although there have been many content replacement strategies, TTL is selected because: i) this paper mainly focuses on the inter-slice level resource allocation, and it is reasonable to apply a well-known caching method for intra-slice caching; ii) TTL requires no prior knowledge of content popularity, which is more realistic. Nevertheless, our proposed architecture is compatible with any other caching methods without loss of generality. The complexity of the slicing-based caching model lies in how to effectively divide the limited caching capacity between slices, which is far more complicated than the original TTL model. We use $\mathcal{N}_{j}$ to represent the slice set in the BS $j$, and the slice $n$ contains $|\mathcal{M}_{j,n}|$ UEs, and each UE is denoted by $m$. Each slice has its own content catalog $\mathcal{G}_{j,n}$, and the variables $g$ represents the content items ($g\in \mathcal{G}_{j,n}$). We assume all content items have the same packet size \cite{b25-2}. 

$\phi_{j,n,m,g}$ represents the request rate of UE $m$ for the content item $g$, which denotes the frequency that content item $g$ is demanded by the UE $m$. Then we have
\begin{equation} \label{eq3-0}
\phi_{j,n,m,g}=p_{j,n,m,g}\phi_{j,n,m},
\end{equation}
where $\phi_{j,n,m}$ is the total request rate of UE $m$, and $p_{j,n,m,g}$ is the request rate distribution of UE $m$ for content items ($\sum_{g\in\mathcal{G}} p_{j,n,m,g}=1 $). 

The cache hit ratio indicates the probability of finding a content in the cache, which can be calculated by \cite{b25}:
\begin{equation} \label{eq3}
h_{j,n,g}=1-e^{-\sum\limits_{m\in \mathcal{M}_{j,n}}\phi_{j,n,m,g}T_{j,n,m,g}},
\end{equation}
where $h_{j,n,g}$ is the cache hit ratio of content item $g$ in the slice $n$, and the TTL of the content item $g$ will be reset to $T_{j,n,m,g}$ when this content item is required. Note that the number of contents is large and the request rate of each content item is relatively small. Based on $\lim\limits_{x\to 0}e^{x}=x+1$, we approximately have:
\usetagform{blue}
\begin{equation} \label{eq4}
h_{j,n,g}=\sum_{m\in \mathcal{M}_{j,n}}\phi_{j,n,m,g}T_{j,n,m,g}.
\end{equation}

Meanwhile, the total cache hit ratio is related to the allocated storage capacity of slice $n$: 
\usetagform{default} 
\begin{equation} \label{eq5}
\sum _{g\in \mathcal{G}_{n}}\sum _{m\in \mathcal{M}_{n}}h_{j,n,m,g}=\frac{C_{j,n}}{C_{j,T}},
\end{equation}
where $C_{j,n}$ is the allocated storage capacity for slice $n$ in BS $j$, $C_{j,T}$ is the total storage capacity of BS $j$, and $h_{j,n,m,g}$ is the cache hit ratio of UE $m$ for content item $g$. We assume the content item $g$ has the same popularity for different UEs and thus $T_{j,n,m,g}=T_{j,n,g}$. Given equation (\ref{eq4}) and (\ref{eq5}), we have (the proof is given in the appendix):
 
\begin{equation} \label{eq6}
h_{j,n,g}=\frac{C_{j,n}}{C_{j,T}}\frac{\sum_{m\in \mathcal{M}_{j,n}}\phi_{j,n,m,g}}{\sum_{m\in \mathcal{M}_{j,n}}\phi_{j,n,m}}.
\end{equation}     

Equation (\ref{eq6}) indicates that a higher caching capacity $C_{n}$ leads to a higher cache hit ratio $h_{n,g}$ \cite{b25-2}. As such, popular content items, which is indicated by a higher request rate $\phi_{n,m,g}$, has a higher cache hit ratio.

Then we define a cache hit rate $\phi_{j,n,m}^{hit}$ to describe the frequency of requesting cached content items in the total request rate, which can be calculated by: 
\begin{equation} \label{eq6-1}
\phi_{j,n,m}^{hit}=\sum _{g\in \mathcal{G}_{j,n}}\phi_{j,n,m,g}h_{j,n,m,g},
\end{equation} 
and the cache miss rate is:
\begin{equation} \label{eq6-11}
\phi_{j,n,m}^{miss}=\phi_{j,n,m}-\phi_{j,n,m}^{hit},
\end{equation} 
which indicates the frequency of requesting non-cached content items in the total request rate. 

Finally, backhaul delay is only applied when a content item is not cached at the BS. The backhaul service is presumed to obey the M/M/1 queue, and the average delay is\cite{b26}:
\begin{equation} \label{eq7}
d^{back}_{j,m,g}=\frac{1}{\frac{B}{L_{m}}-\sum_{n\in \mathcal{N}}\sum_{m\in \mathcal{M}_{j,n}}(\phi_{j,n,m}^{miss})},
\end{equation}
where $B$ is the backhaul capacity, $L_{m}$ is the average size of content items, $B/{L_{m}}$ denotes the service rate, and $\sum_{n\in \mathcal{N}}\sum_{m\in \mathcal{M}_{j,n}}(\phi_{j,n,m}^{miss})$ is the total cache miss rate. Note that backhaul delay will be infinite if the cache miss rate is higher than the service rate.

\begin{table*}[!t]
\caption{Summary of strategies and MDP definitions}
\centering
\renewcommand\arraystretch{2}
 \begin{threeparttable}  
\begin{tabular}{|m{1.2cm}<{\centering}|m{5.8cm}<{\centering}|m{1.7cm}<{\centering}|m{3.8cm}<{\centering}|m{3.8cm}<{\centering}|}
\hline
Indices & Strategies & States & Actions\tnote{2} & Instant Reward\\
\hline
Expert 1 & Q-learning based radio resource allocation; No caching capability. & 
\multirow{5}*{\makecell[c] {\quad\\\quad\\ \quad\\ $(s^{embb},s^{urllc})$, \\where $s^{embb}$ \\denotes the \\ number of \\ eMBB packets \\ in the queue, \\ and $s^{urllc}$ \\is defined \\similarly.}}  & $(N^{embb},N^{urllc})$  & 
\multirow{6}*{\makecell[l] { $r=wr^{embb}+(1-w)r^{urllc}$, 
\\where\\ 
$ r^{embb}=\frac{2}{\pi} tan^{-1}(b^{embb,avg})$\\
$ r^{urllc} = \frac{2}{\pi} tan^{-1}(d^{tar}-$\\$d^{urllc,avg})$. $w$ is the weighting \\factor, $r^{embb}$ and $r^{urllc}$ are \\ the rewards of eMBB and \\ URLLC slices, respectively.\\ $b^{embb,avg}$ and $d^{urllc,avg}$\\ are average throughput \\ of eMBB slice and average \\ delay of URLLC slice. $d^{tar}$ is \\ the target URLLC delay.} } \\
\cline{1-2}
\cline{4-4}
Expert 2 & Fixed radio resource allocation; Q-learning based caching capacity allocation.  & ~  & $(C^{embb},C^{urllc})$  &~ \\
\cline{1-2}
\cline{4-4}
Learner 1 (QDTRL) & DTRL-based joint radio and cache resource allocation with Q-values mapping function.  & ~  & \multirow{3}*{\makecell[c] { \quad\\$(N^{embb},N^{urllc}$,$C^{embb},C^{urllc})$,\\ where $N^{embb}$ and $C^{embb}$\\ denote the number of RBs and\\ caching capacity allocated to \\ the eMBB slice, respectively.\\ $N^{urllc}$ and $C^{urllc}$ are defined \\ similarly for the URLLC slice.}}  &~ \\
\cline{1-2}
Learner 2 (ADTRL) & DTRL-based joint radio and  cache resource allocation with action selection mapping function.  & ~  &~  &~ \\
\cline{1-2}
Baseline\tnote{1}  & EB-DQN for joint radio and cache resource allocation without any prior knowledge.  & ~  & ~ &~\\
\hline
\end{tabular}

 \begin{tablenotes}    
        \footnotesize       
        \item[1] We include the PPF-TTL as the second baseline. However, since PPF-TTL is not an ML-based method, it is excluded from this table. 
        \item[2] Given the total number of RBs $N_{j,T}$, $N^{urllc}$ can be easily calculated if $N^{embb}$ has been decided. However, we present the action definition using $(N^{embb},N^{urllc})$ for better readability and scalability if more slices are included, and similarly for the action definition of cache resource allocation $(C^{embb},C^{urllc})$
      \end{tablenotes}            

\end{threeparttable}  
\label{tab1}
\vspace{-10pt}
\end{table*}

\subsection{Problem Formulation}
The objective of the eMBB slice is to maximize the total throughput, while the URLLC slice aims to minimize the average delay. It is assumed that the content catalogs of the two slices are not overlapped. To balance the requirements of the two slices, the GRM needs to jointly consider the objectives of the two slices and allocate the radio and cache resource accordingly. For each BS, the GRM allocates radio and cache resource by: 
\begin{subequations}\label{eq8:main}
\begin{align}
\text{max}  \quad & wb^{embb,avg}+(1-w)(d^{tar}-d^{urllc,avg}),& \tag{\ref{eq8:main}} \\
 \text{s.t.} \quad & (\ref{eq1}) \, (\ref{eq2})\, (\ref{eq6})\, (\ref{eq6-1})\, \text{and}\,(\ref{eq7})  \\
             & \sum_{n\in\mathcal{N}_{j}}N_{j,n}\leq N_{j,T} & \label{eq8-2}\\
             & \sum_{n\in\mathcal{N}_{j}}C_{j,n}\leq C_{j,T} & \label{eq8-3}\\
             & \sum_{n\in\mathcal{N}_{j}}\sum_{m\in \mathcal{M}_{j,n}}\phi_{m}^{miss}\leq\frac{B}{L_{m}} & \label{eq8-4}\\
             & \sum_{n\in\mathcal{N}_{j}} \sum_{m\in \mathcal{M}_{j,n}}x_{j,q,m}\leq1 & \label{eq8-5}  \\
             & \sum_{m\in \mathcal{M}_{j,n}}\sum_{q\in \mathcal{N}^{RB}_{j}}x_{j,q,m}\leq N_{j,n} & \label{eq8-6}  \\
             & \sum_{g\in \mathcal{G}_{j,n}} \beta_{j,g} \leq C_{j,n} & \label{eq8-7}
\end{align}
\end{subequations}
where $b^{embb,avg}$ is the total throughput of the eMBB slice, $d^{urllc,avg}$ is the average latency of the URLLC slice, and $d^{tar}$ is the target delay. Here we use $w$ as a weight factor in (\ref{eq8:main}) to balance the objectives of the two slices and maximize the overall objective. $\mathcal{N}_{j}$ is the slice set of BS $j$, which consists of eMBB and URLLC slices. $N_{j,n}$ is the number of RBs that the GRM allocated to the slice $n$, and $N_{j,T}$ is the total number of RBs of BS $j$, and the equation (\ref{eq8-2}) is the radio resource constraint. $C_{j,n}$ is the caching capacity of slice $n$, and $C_{j,T}$ is the total caching capacity in BS $j$, and (\ref{eq8-3}) ensures that the allocated caching capacity cannot exceed the upper limit. $\mathcal{M}_{j,n}$ is the set of UEs in slice $n$, and $\phi_{m}^{miss}$ is the miss rate of UE $m$, and (\ref{eq8-4}) denotes that the total miss rate should not exceed the backhaul service rate. $x_{j,q,m}$ has been defined in equation (\ref{eq2}) as the RB allocation indicator. Equations (\ref{eq8-5}) and (\ref{eq8-6}) denote that one RB can only be allocated to one UE, and the total number of available RBs are $N_{j,n}$ in slice $n$. Finally, $\beta_{j,g}$ is a binary variable that has been defined in equation (\ref{eq1}) to represent whether a content item is cached.      

In the defined problem formulations, $N_{j,n}$ and $C_{j,n}$ are the control variables of the GRM, which means the GRM only accounts for the inter-slice resource allocation. $x_{j,q,m}$ and $\beta_{j,g}$ are the control variables of SRMs. In SRMs, we apply classic proportional fairness algorithm for intra-slice RB allocation to determine $x_{j,q,m}$, since all UEs in the same slice are presumed to be equally important\cite{b20-0}. Meanwhile, cached content items are updated by TTL rule, which will determine $\beta_{j,g}$.

\section{Deep Transfer Reinforcement Learning based Resource Allocation}
\label{s4}
\subsection{Overall framework}
In this section, we introduce the DTRL-based inter-slice resource allocation, where each BS is considered as an independent agent to make decisions autonomously. As shown in Table \ref{tab1}, five learning-based strategies are deployed. We assume experts 1 and 2 apply Q-learning for radio and cache resource allocation, respectively. Experts are only good at one of radio or cache resource allocation, but they have no multi-task knowledge. Then learners 1 and 2 can utilize knowledge from experts to improve their own performance on joint radio and cache resource allocation. Based on different mapping functions, we propose two DTRL-based methods, namely QDTRL and ADTRL. Finally, we apply EB-DQN as a learning-based benchmark and the PPF-TTL method as a model-based baseline. In the following, we will introduce the experts, learners and baselines.  

\subsection{Q-learning based Experts}
In this section, we assume the expert agents have learning experience on one specific task, but they have no knowledge of other tasks. For expert 1, it uses Q-learning for the radio resource allocation, and there is no caching capability. For expert 2, RBs are allocated by the numbers of UEs in each slice, and Q-learning is used for the caching capacity allocation. 

To transform the problem formulation in equation (\ref{eq8:main}) to the RL context, we first define the MDP $(S,A,T,R)$ for experts, where $S$ is the state set, $A$ is the action set, $T$ is the transition probability, and $R$ is the reward function. The MDP definitions of experts are given below:  
\begin{itemize}
  \item \textbf{State}: In this work, we intend to coordinate the performance of various slices by inter-slice resource allocation. As such, the state definition should reflect the transmission demand of each slice. The states of expert 1 and 2 are both defined by $(s^{embb},s^{urllc})$, which indicates the number of packets waiting in the queues of eMBB and URLLC slices, respectively. 
  \item \textbf{Action of expert 1}: The expert 1 only implements radio resource allocation, and consequently the action $(N^{embb},N^{urllc})$ denotes the number of RBs allocated to eMBB and URLLC slices. 
  \item \textbf{Action of expert 2}: In expert 2, the learning strategy is only applied for caching capacity allocation, and the action $(C^{embb},C^{urllc})$ indicates the caching capacity allocated to eMBB and URLLC slices. \\
  It is worth noting that the actions $N^{embb},N^{urllc}$, $C^{embb}$ and $C^{urllc}$ have been defined as control variables in the problem formulation (\ref{eq8:main}), and they are transformed to actions here to serve the Q-learning scheme.
  \item \textbf{Reward function}: The reward functions of expert 1 and 2 are defined by the objectives of slices:
  \begin{equation} \label{e9}
  r=wr^{embb}+(1-w)r^{urllc},
  \end{equation}
  where $r^{embb}$ and $r^{urllc}$ are rewards of eMBB and URLLC slices, respectively, and $w$ is the weight factor.\\
  For the eMBB slice, obtaining higher throughput leads to a higher reward, and we have.
      \usetagform{blue}
      \begin{equation} \label{e10}
      r^{embb}=\frac{2}{\pi} tan^{-1}(b^{embb,avg}),
      \end{equation}
   where $b^{embb,avg}$ is the average throughput of the eMBB slice, and we apply the $tan^{-1}$ function to normalize the reward ($0<r^{embb}<1$).  \\
   For the URLLC slice, lower delay means a higher reward:
   \usetagform{default}
 \begin{equation} \label{e11}
r^{urllc} = \frac{2}{\pi} tan^{-1}(d^{tar}-d^{urllc,avg}),
\end{equation}
where $d^{tar}$ and $d^{urllc,avg}$ are target and achieved average delays for the URLLC slice, respectively. Note that both $r^{embb}$ and $r^{urllc}$ are normalized to balance the performance metrics of the two slices.\\
Moreover, to guarantee the constraints of the problem formulation (\ref{eq8:main}), we apply penalties to the reward when these constraints are violated.
\end{itemize}

With Q-learning, the agent aims to maximize the long-term expected reward:
\begin{equation} \label{eq17}
V(s_{e}) =\mathbb{E}_{\pi}(\sum_{i=0}^{\infty}\gamma^{i} r(s_{e,i},a_{e,i})|s_{e}=s_{e,0}),
\end{equation}
where $V(s_{e})$ is the long-term expected accumulated reward of experts at state $s_{e}$, and here we use notation $e$ to indicate experts. $s_{e,0}$ is the initial state, and $r(s_{e,i},a_{e,i})$ denotes the reward of selecting action $a_{e,i}$ at state $s_{e,i}$ in episode $i$, and $\gamma$ is the discount factor $(0<\gamma<1)$.

Then the state-action values of expert 1 and 2 are updated by:
\usetagform{blue}
\begin{equation} \label{eq18}
\begin{aligned}
Q^{new}(s_{e},a_{e})&= Q^{old}(s_{e},a_{e})+\\&\alpha(r_{e}+\gamma \max\limits_{a} Q(s'_{e},a) -Q^{old}(s_{e},a_{e})),
\end{aligned}
\end{equation}
where $Q^{old}(s_{e},a_{e})$ and $Q^{new}(s_{e},a_{e})$ are old and new Q-values, $s_{e}$ and $s_{e}'$ are current and next states of experts, respectively, $a_{e}$ is the action, $r_{e}$ is the reward, and $\alpha$ is the learning rate ($0< \alpha < 1 $). By updating the Q-values iteratively, experts will learn the optimal action selections to achieve the best accumulated reward.  

\subsection{Deep Transfer Reinforcement Learning based Learners}
In this section, we propose two DTRL-based algorithms, namely QDTRL (learner 1) and ADTRL (learner 2). QDTRL utilizes the Q-values of the experts as prior knowledge, while ADTRL uses the action selection experience of experts for improved performance. Consequently, we define two different mapping functions to transfer knowledge from experts to learners 1 and 2, respectively.

In this section, we first define the states, actions and rewards of learners, then we introduce the DDQN framework for two learners. Finally, we present the proposed QDTRL and ADTRL algorithms. 

\textit{1) MDP and DDQN Framework for Learners}

Given the prior knowledge of experts, learners are expected to solve more complicated problems with a larger state-action space. To apply TRL, we define the MDP by $(S,A,R,T,\mathcal{F})$, where $\mathcal{F}$ is the mapping function. The states, actions and reward functions of the two learners are given below: 
\begin{itemize}
  \item \textbf{State and reward function}: As shown in Table \ref{tab1}, we assume learners have the same state and reward definitions as experts. The main reason is that transfer learning is designed for tasks that share some similarities with existing expert tasks. As such, similar state and reward definitions can reduce the complexity of defining mapping functions, which can better transfer the knowledge from experts to learners.  
  \item \textbf{Action of learners 1 and 2}: Compared with experts, the learners have to jointly consider the radio and cache resource allocation, and then the actions are defined by ($N^{embb},N^{urllc},C^{embb},C^{urllc}$). Compared with allocating one single resource, the joint resource allocation problem is much more complicated, especially when multiple slices are involved.
\end{itemize}

Based on MDP definitions, we introduce our DTRL algorithm. 
In conventional Q-learning, Q-values are updated by:
\begin{equation} \label{eq18-1}
\begin{aligned}
Q^{new}(s_{l},a_{l})&= Q^{old}(s_{l},a_{l})+\\&\alpha(r_{l}+\gamma \max\limits_{a} Q(s'_{l},a) -Q^{old}(s_{l},a_{l})),
\end{aligned}
\end{equation}
where $s_{l}$ and $a_{l}$ are the state and action of the learner, respectively, $s_{l}'$ is the next state, and $r_{l}$ is the reward. Here we use the notation $l$ to indicate the learner. Q-learning applies a Q-table to record state-action values, and consequently it may suffer a slow convergence issue when the state-action space is huge. To this end, DQN is proposed by using deep neural networks to predict Q-values \cite{b22}. 

When the Q-values in equation (\ref{eq18-1}) converge, we have $Q^{old}(s_{l},a_{l})$ $=Q^{new}(s_{l},a_{l})$ and $Q^{old}(s_{l},a_{l})=r_{l}+\gamma \max\limits_{a} Q(s_{l}',a)$. Then, a loss function can be defined for the network training of DQN:
\usetagform{blue}
\begin{equation} \label{eq18-11}
L(w)=Er(r_{l}+\gamma \max\limits_{a} Q(s_{l}',a,w')-Q(s_{l},a_{l},w)),
\end{equation}
where $Er$ is the loss function to represent the error between prediction results $r_{l}+\gamma \max\limits_{a} Q(s_{l}',a,w')$ and target results $Q(s_{l},a_{l},w)$. $w$ is the weight of the main network, which will predict current Q-values $Q(s_{l},a_{l},w)$. $w'$ is the weight of the target network, and it predicts the target Q-values $Q(s_{l}',a,w')$. 

\begin{figure*}[!t]
\centering
\vspace{-10pt}
\includegraphics[width=17.5cm,height=6.5cm]{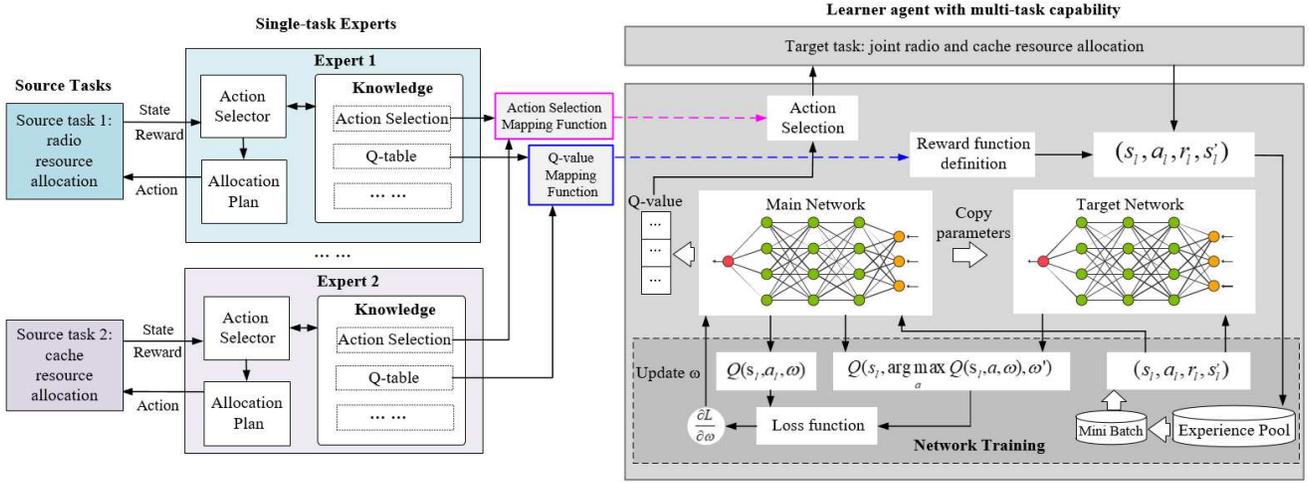}
\caption{Proposed deep transfer reinforcement learning architecture for resource management.}
\label{fig1-2}
\vspace{0pt}
\end{figure*}

In DQN, note that the action selection and evaluation are both implemented by the target network, which is indicated by $\max\limits_{a} Q(s_{l}',a,w')$. Meanwhile, target Q-values are calculated by the maximum Q-value of the next state. If the maximize operator is always included in the Q-value calculation, then the Q-value predicted by neural networks will be obviously higher every time \cite{b26-02}. To this end, the DDQN has been proposed to decouple the action selection and evaluation. The loss function of DDQN is defined as:
\begin{equation}
\label{eq18-2}
\begin{aligned}
L&(w)=Er(r_{l}+ \\
&\gamma Q(s_{l}',\arg \max\limits_{a}Q(s_{l}',a,w),w') - Q(s_{l},a_{l},w)),
\end{aligned}
\end{equation}
where the main network chooses actions by $a_{l}=\arg \max\limits_{a}Q(s_{l}',a,w))$, and the target network evaluates the action by $Q(s_{l}',a_{l},w')$. By decoupling the action selection and evaluation, DDQN can prevent overestimation and better predict Q-values than DQN.

In this work, we deploy the DDQN architecture in the proposed DTRL. For the learner agent shown by color grey in Fig.\ref{fig1-2}, an action $a_{l}$ is first selected and sent to the environment. Then a tuple $(s_{l},a_{l},r_{l},s_{l}')$ will be received from the environment, which will be saved in the experience pool. The learner agent samples a random minibatch from the experience pool. For every tuple $(s_{l},a_{l},r_{l},s_{l}')$, the main network predicts $Q(s_{l},a_{l},w)$ and selects actions by $a=\arg \max\limits_{a}Q(s_{l}',a,w))$. The target network evaluates the action by $Q(s_{l}',a,w')$. Then we utilize the loss function shown by equation (\ref{eq18-2}) for gradient descent to update the weight $w$ of the main network. After several training sessions, the target network will copy the weight parameters of the main network. Such a late update of the target network serves as a stable reference for the main network training. Here we deploy the Long Short-Term Memory (LSTM) network as hidden layers for main and target networks. As a special recurrent neural network, LSTM can better capture the long term data dependency, which makes it an ideal candidate to handle complicated wireless network environments\cite{b28-00}. 

Finally, it is worth noting that we include two different mapping functions in the proposed DTRL scheme. The Q-value mapping function will affect the reward calculation of the learner agent (indicated by the pink line in Fig.\ref{fig1-2}), while the action selection mapping function influences the action selection (shown by the blue line in Fig.\ref{fig1-2}). Accordingly, we propose two DTRL-based methods, namely QDTRL and ADTRL, and in the following we will introduce these two mapping functions and corresponding algorithms.

\textit{ 2) Learner 1: Q-value based Deep Transfer Reinforcement Learning}

In QDTRL, the Q-values of the experts are presumed to be the prior knowledge of the learner. The main idea behind this is to encourage learners to select actions that have higher Q-values in the experts. Considering the task similarities, actions with a higher Q-value of experts are very likely to bring similar high rewards for the learner. In particular, we consider the Q-values of the experts as extra rewards for learners, which is expected to improve exploration efficiency by selecting actions with higher potential rewards\cite{b26-0}.

Firstly, the loss function in QDTRL is defined by: 
\begin{equation}
\label{eq19}
\begin{aligned}
L&(w)=Er(\sigma_{1} Q^{E}(\mathcal{F}(s_{l}),\mathcal{F'}(a_{l}))+r_{l}+ \\
&\gamma Q(s_{l}',\arg \max\limits_{a}Q(s_{l}',a,w),w') - Q(s_{l},a_{l},w)),
\end{aligned}
\end{equation}
where $\mathcal{F}$ and $\mathcal{F'}$ are state and action mapping functions, respectively. Compared with equation (\ref{eq18-2}), the main difference is that $\sigma_{1} Q^{E}(\mathcal{F}(s_{l}),\mathcal{F'}(a_{l}))$ is involved as an extra reward of selecting action $a_{l}$ under state $s_{l}$.  $\sigma_{1}$ is the transfer learning rate, which describes the importance of prior knowledge ($0\leq \sigma_{1} \leq 1 $).
A higher transfer learning rate means the prior knowledge utilization is more important than its own learning process, while a lower value indicates the reverse. 

In equation (\ref{eq19}), we apply $\sigma_{1} Q^{E}(\mathcal{F}(s_{l}),\mathcal{F'}(a_{l}))$ to guide the action selection of learner. However, due to different state-action spaces, the Q-values of the experts cannot be directly utilized by the learner, thus a function is needed to map experts' Q-values to the learner's Q-table. The Q-value mapping function consists of state mapping and action mapping, and $Q^{E}$ term in equation (\ref{eq19}) is generated by: 
\usetagform{default}
\begin{equation} \label{eq20}
\resizebox{0.89\hsize}{!}{$\begin{aligned}
Q^{E}(\mathcal{F}(s_{l}),\mathcal{F'}(a_{l}))= Q_{e,1}(s_{e,1},a_{e,1})+Q_{e,2}(s_{e,2},a_{e,2}), 
\end{aligned}$}
\end{equation}
where $Q_{e,1}$, $s_{e,1}$ and $a_{e,1}$ are the Q-value, state and action of expert 1, respectively. $Q_{e,2}$, $s_{e,2}$ and $a_{e,2}$ are defined similarly for expert 2. The objective of mapping functions $\mathcal{F}$ and $\mathcal{F'}$ is to find the states and actions of the experts that are close to $s_{l}$ and $a_{l}$. Considering the task similarities, we can use the existing decision knowledge of the expert agent to guide the action selection of learners by finding similar states and actions \cite{b26-01}. $\mathcal{F}$ and $\mathcal{F'}$ are defined by:  
\begin{itemize}
  \item \textbf{State mapping $\mathcal{F}$}: For a given state $s_{l}$, considering experts and learner 1 have the same state definition, we can always find $s_{l}=s_{e,1}=s_{e,2}$. Thus $\mathcal{F}$ can be easily defined.
  \item \textbf{Action mapping $\mathcal{F'}$}: The goal of $\mathcal{F'}$ is to find  $(a_{e,1},a_{e,2})=\mathcal{F'}(a_{l})$.  For any action $a_{l}$, which is defined as $a_{l}=(N^{embb},N^{urllc}$,$C^{embb},C^{urllc})$, it can be decomposed into the combination of $a_{e,1}=(N^{embb},N^{urllc})$ and $a_{e,2}=(C^{embb},C^{urllc})$. Then $\mathcal{F'}$ can be defined accordingly. 
\end{itemize}

\begin{figure*}[!t]
\centering
\vspace{-10pt}
\includegraphics[width=11cm,height=5.5cm]{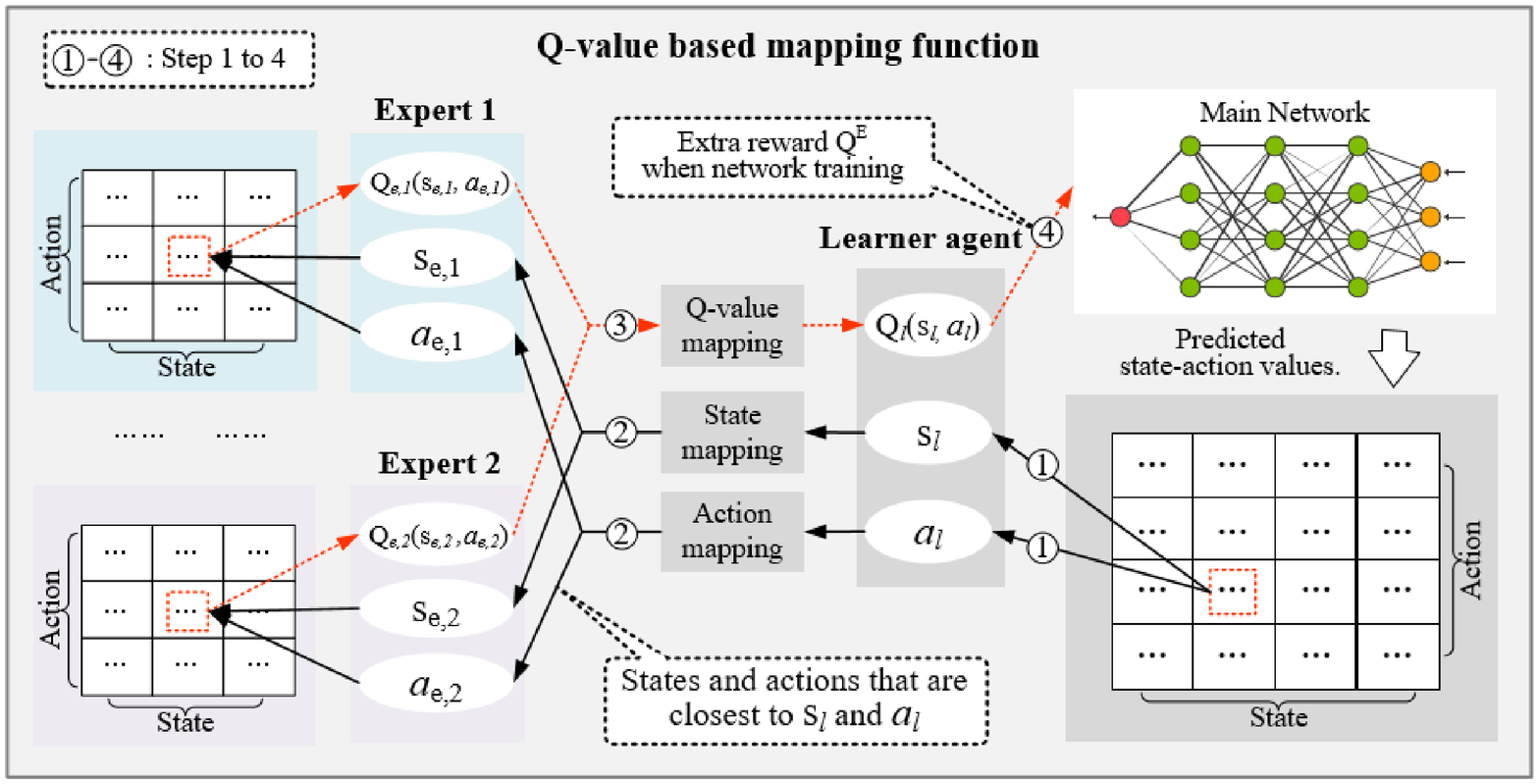}
\caption{Proposed Q-value-based mapping function for deep transfer reinforcement learning.}
\label{fig1-4}
\vspace{-10pt}
\end{figure*}

Based on the state and action mapping relationships, if given $s_{l}$ and $a_{l}$, we can always find specific $Q_{e,1}(s_{e,1},a_{e,1})$ and $Q_{e,2}(s_{e,2},a_{e,2})$. Then $Q^{E}(\mathcal{F}(s_{l}),\mathcal{F'}(a_{l}))$ can be directly used by learner 1.
The defined Q-value mapping function is summarized by Fig.\ref{fig1-4} from step 1 to 4. First, for any given $(s_{l},a_{l})$, we find state action pairs $(s_{e,1},a_{e,1})$ and $(s_{e,2},a_{e,2})$ by state mapping function $\mathcal{F}$ and action mapping function $\mathcal{F'}$. Then, we extract $Q_{e,1}(s_{e,1},a_{e,1})$ and $Q_{e,2}(s_{e,2},a_{e,2})$ from expert agents' Q-tables. After that, we generate $Q^{E}(\mathcal{F}(s_{l}),\mathcal{F'}(a_{l}))$ by equation (\ref{eq20}), which will be considered as extra rewards when selecting $a_{l}$ under $s_{l}$. This extra reward is added to $r_{l}$, and the new tuple $(s_{l},a_{l},r_{l}+\sigma Q^{E}(\mathcal{F}(s_{l}),\mathcal{F'}(a_{l})),s_{l}')$ will be saved in the experience pool. Finally, we implement the gradient descent by equation (\ref{eq19}) for the network training.

\textit{3) Learner 2: Action Selection based Deep Transfer Reinforcement Learning}

Learners are mainly designed to handle more complicated problems than experts, which usually means larger state or action spaces. For instance, the joint resource allocation problem has higher action spaces than allocating one single resource, result a very large action space and longer convergence. To this end, we propose an ADTRL algorithm to improve exploration efficiency by evaluating the potential optimality of actions. 
More specifically, we first apply a lax bisimulation metric to assess the MDP similarities between learners and experts. Then, we calculate the potential advantage of different actions in the learner and produce a lower bound for the optimality. Finally, the optimality metrics of different actions are normalized, and we assume that actions with higher potential optimality are more likely to be selected in the exploration phase. In the following, we will introduce the proposed method in detail.

It is worth noting that TL is mainly applied to learner tasks that are related to expert tasks. Specifically, it requires similarities between expert and learner MDPs. Then, we first introduce the Kantorovich distance $\mathcal{K}(\mathcal{D})(Y,Z)$ to describe the similarities between two distributions\cite{b26-1}:
\begin{equation}\label{eq21}
\begin{aligned}
\max\limits_{u_{f},f=1,2,...,|S|}  \quad & \sum_{f=1}^{|S|}(Y(s_{f})-Z(s_{f}))u_{f},&  \\
 \text{s.t.} \quad & u_{f}-u_{k} \leq \mathcal{D}(s_{f},s_{k}) \quad f,k=1,2,...,|S| \\
             & 0 \leq u_{f} \leq 1 
\end{aligned}  
\end{equation}
where $Y$ and $Z$ are two probability distributions of $s_{f} \in S$, $u_{f}$ and $u_{k}$ are internal optimization variables, and $\mathcal{D}(s_{f},s_{k})$ denotes a metric $\mathcal{D}$ to assess the distance between $s_{f}$ and $s_{k}$. $\mathcal{K}(\mathcal{D})(Y,Z)$ shows the distance of two probability distributions $Y$ and $Z$ with the metric $\mathcal{D}$ in set $S$. 
However, in TL, we focus more on the distance metric of different MDPs, and $\mathcal{K}(\mathcal{D})(Y,Z)$ is rewritten by\cite{b26-11}:
\begin{equation}\label{eq22}
\begin{aligned}
\max\limits_{u_{f},f=1,2,...,|S_{1}|;v_{k},k=1,2,...,|S_{2}|} & \quad  \sum_{f=1}^{|S_{1}|}Y(s_{f})u_{f}- \sum_{k=1}^{|S_{2}|}Z(s_{k})v_{k},&  \\
 \text{s.t.} \quad & u_{f}-v_{k} \leq \mathcal{D}(s_{f},s_{k})\\
             & -1 \leq u_{f} \leq 1 
\end{aligned} 
\end{equation}
where $S_{1}$ and $S_{2}$ are two sets with $s_{f}\in S_{1}$ and $s_{k}\in S_{2}$, respectively, and $\mathcal{K}(\mathcal{D})(Y,Z)$ evaluates the distance between two distribution set $S_{1}$ and $S_{2}$. Here $f=1,2,..,|S_1|$ means that set $s_f$ has $|S_1|$ possible values in set $S_1$, and the probability distribution function $Y$ of $s_f$ satisfies $\sum_{f=1}^{|S_1|}Y(s_{f})=1$. $s_k$, $S_2$ and the probability distribution function $Z$ can be defined similarly.

Then we include two MDPs $<S_{e},A_{e},T_{e},R_{e}>$ and $<S_{l},A_{l},$ $T_{l},R_{l}>$ to identify the difference of their state-action pairs. The lax bisimulation metric is introduced to evaluate the distance between two state-action pairs\cite{b26-2}:
\begin{equation}
\begin{aligned}\label{eq23}
\mathcal{D}_{\sim}((s_{e},a_{e}),(s_{l},a_{l}))&:=\theta_{1}|r_{e}(s_{e},a_{e})-r_{l}(s_{l},a_{l})|\\
&+\theta_{2} \mathcal{K}(\mathcal{D'}(Y(s_{e},a_{e}),Z(s_{l},a_{l}))),  
\end{aligned}
\end{equation}           
where $\theta_{1}$ and $\theta_{2}$ are weight factors. The first term $|r_{e}(s_{e},a_{e})-r_{l}(s_{l},a_{l})|$ represents the reward distance, and $\mathcal{K}(\mathcal{D'}(Y(s_{e},a_{e}),Z(s_{l},a_{l})))$ is the Kantorovich metric for state-action pairs under semi-metric $\mathcal{D'}$. $\mathcal{D'}$ is defined by the Hausdorff metric: 
\begin{equation}\label{eq24}
\begin{aligned}
\mathcal{D'}(s_{e},s_{l})=\max(\max\limits_{a_{e}\in A_{e}}& \min\limits_{a_{l}\in A_{l}}\mathcal{D}((s_{e},a_{e}),(s_{l},a_{l})),\\
&\min\limits_{a_{e}\in A_{e}}\max\limits_{a_{l}\in A_{l}}(\mathcal{D}(s_{e},a_{e}),(s_{l},a_{l}))). 
\end{aligned}
\end{equation}
Hausdorff distance is the maximum distance from one set to the nearest point of the other set \cite{b26-3}, and here we define $\mathcal{D'}(s_{e},s_{l})$ to measure the distance between action sets $A_{e}$ and $A_{l}$ under state $s_{e}$ and $s_{l}$.

To evaluate the potential optimality of selecting $a_{l}$ under $s_{l}$, we include the Bellman optimality to find the state value difference:
\begin{equation}\label{eq25}
\resizebox{0.8\hsize}{!}{$\begin{aligned}
&|Q^{*}_{l}(s_{l},a_{l})-V_{e}^{*}(s_{e})|\\
&=|Q^{*}_{l}(s_{l},a_{l})-Q^{*}_{e}(s_{e},\pi^{*}(s_{e}))|\\
&=|Q^{*}_{l}(s_{l},a_{l})-Q^{*}_{e}(s_{e},a^{*}_{e})|\\
&=|(r_{l}(s_{l},a_{l})+\gamma\sum\limits_{s_{l}'\in S_{l}}Y(s_{l}'|s_{l},a_{l})V_{l}(s_{l}'))- \\
&\qquad \qquad (r_{e}(s_{e},a^{*}_{e})+\gamma\sum\limits_{s_{e}'\in S_{e}}Z(s_{e}'|s_{e},a^{*}_{e})V_{e}(s_{e}'))| \\
&\leq|r_{l}(s_{l},a_{l})-r_{e}(s_{e},a^{*}_{e})|+\\
&\qquad \gamma|\sum\limits_{s_{l}'\in S_{l}}Y(s_{l}'|s_{l},a_{l})V_{l}(s')-\sum\limits_{s_{e}'\in S_{e}}Z(s_{e}'|s_{e},a^{*}_{e})V_{e}(s'))|\\
&\leq|r_{l}(s_{l},a_{l})-r_{e}(s_{e},a^{*}_{e})|+\max\limits_{a_{l}\in A_{l}}\min\limits_{a_{e}\in A_{e}}(\\
&\qquad \gamma|\sum\limits_{s_{l}'\in S_{l}}Y(s_{l}'|s_{l},a_{l})V_{l}(s')-\sum\limits_{s_{e}'\in S_{e}}Z(s_{e}'|s_{e},a^{*}_{e})V_{e}(s')|)\\
&\qquad \qquad \text{\qquad (Using equation (\ref{eq22}), (\ref{eq24}))}\\
&=|r_{e}(s_{e},a^{*}_{e})-r_{l}(s_{l},a_{l})|+\gamma \mathcal{K}(\mathcal{D'}((s_{e},a^{*}_{e}),(s_{l},a_{l})))\\
&\text{\qquad \qquad (Setting $\theta_{1}=1$ and $\theta_{2}=\gamma$ in equation (\ref{eq23}))}\\
&= \mathcal{D}_{\sim}((s_{l},a_{l}),(s_{e},a^{*}_{e})),
\end{aligned}$}
\end{equation}
where $Q^{*}_{l}(s_{l},a_{l})$ is the optimal state-action value of $(s_{l},a_{l})$, $V^{*}_{e}(s_{e})$ is the optimal state value of $s_{e}$, $s'_{l}$ is the next state of $s_{l}$, and $Y(s'|s_{l},a_{l})$ is the probability of arriving to $s'_{l}$ by implementing $a_{l}$ under $s_{l}$. Here we use $a_{e}^{*}=\pi^{*}(s_{e})$ to represent the action selection of the expert agent. Equation (\ref{eq25}) shows that there is a upper bound between the state-action pairs $(s_{l},a_{l})$ and $(s_{e},a^{*}_{e})$. In the following, we will introduce how to utilize equation (\ref{eq25}) to improve the action selection of learner.

When selecting an action, we usually consider $V^{*}_{l}(s_{l})$ as a target value for $Q_{l}^{*}(s_{l},a_{l})$, and we have $V^{*}_{l}(s_{l})=arg\max\limits_{a_{l}} Q^{*}(s_{l},a_{l})$. Then $V^{*}_{l}(s_{l})-Q_{l}^{*}(s_{l},a_{l})$ can be used to evaluate the potential optimality of $a_{l}$ by:
\begin{equation}\label{eq26}
\resizebox{0.8\hsize}{!}{$\begin{aligned}
&V^{*}_{l}(s_{l})-Q_{l}^{*}(s_{l},a_{l})\\
&=Q^{*}_{l}(s_{l},\pi^{*}(s_{l}))-Q_{l}^{*}(s_{l},a_{l})\\
&=Q^{*}_{l}(s_{l},a^{*}_{l})-Q_{l}^{*}(s_{l},a_{l})\\
&=|Q^{*}_{l}(s_{l},a^{*}_{l})-Q_{l}^{*}(s_{l},a_{l})|\\
&=|Q^{*}_{l}(s_{l},a^{*}_{l})-V_{e}^{*}(s_e)+V^{*}_{e}(s_e)-Q_{l}^{*}(s_{l},a_{l})|\\
&\leq|Q^{*}_{l}(s_{l},a^{*}_{l})-V_{e}^{*}(s_e)|+|V^{*}_{e}(s_e)-Q_{l}^{*}(s_{l},a_{l})|\\
&\leq \mathcal{D}_{\sim}((s_{l},a^{*}_{l}),(s_{e},a^{*}_{e}))+\mathcal{D}_{\sim}((s_{l},a_{l}),(s_{e},a^{*}_{e}))\\
&\text{\quad (By using equation (\ref{eq25})).}
\end{aligned}$}
\end{equation}

Equation (\ref{eq26}) gives an upper bound for the potential optimality of selecting $a_{l}$ under $s_{l}$, and then a lower bound $O_{l}(s_{l},a_{l})$ can be easily found by:
\begin{equation}\label{eq27}
\begin{aligned}
&Q_{l}^{*}(s_{l},a_{l})-V^{*}_{l}(s_{l})\\
&=Q_{l}^{*}(s_{l},a_{l})-Q^{*}_{l}(s_{l},\pi^{*}(s_{l}))\\
&\geq -\mathcal{D}_{\sim}((s_{l},a^{*}_{l}),(s_{e},a^{*}_{e}))-\mathcal{D}_{\sim}((s_{l},a_{l}),(s_{e},a^{*}_{e}))\\
&=O_{l}(s_{l},a_{l}).\\
\end{aligned}
\end{equation}

In $O_{l}(s_{l},a_{l})$, note that $\mathcal{D}_{\sim}((s_{l},a^{*}_{l}),(s_{e},a^{*}_{e}))$ will not affect $a_{l}$ selection, since it is a constant value for given $s_{l}$ and $s_{e}$. Considering we have two experts in this work, we rewrite $O_{l}(s_{l},a_{l})$ by:
\begin{equation}\label{eq27-1}
\resizebox{0.98\hsize}{!}{$\begin{aligned}
&O_{l}(s_{l},a_{l})=-\sigma_{2}(\mathcal{D}_{\sim}((s_{l},a_{l}),(s_{e,1},a^{*}_{e,1}))\\
&+\mathcal{D}_{\sim}((s_{l},a_{l}),(s_{e,2},a^{*}_{e,2})))-(1-\sigma_{2})(\mathcal{D}_{\sim}((s_{l},a^{*}_{l}),(s_{e,1},a^{*}_{e,1}))\\
&+\mathcal{D}_{\sim}((s_{l},a^{*}_{l}),(s_{e,2},a^{*}_{e,2}))),\\
\end{aligned}$}
\end{equation}
where $\sigma_{2}$ is the transfer learning rate of ADTRL (0$\leq\sigma_{2}\leq$1). If $\sigma_{2}=0$, then $O_{l}(s_{l},a_{l})$ becomes a constant value for all $a_{l}\in A_{l}$, which means transferred knowledge will not affect the action selection of the learner. By contrast, if $\sigma_{2}=1$, $O_{l}(s_{l},a_{l})$ totally depends on the lax bisimulation metric between $(s_{l},a_{l})$ and $(s_{e},a^{*}_{e})$, which indicates the learner will imitate the action selection of experts.

\begin{figure}[!t]
\centering
\includegraphics[width=0.85\linewidth]{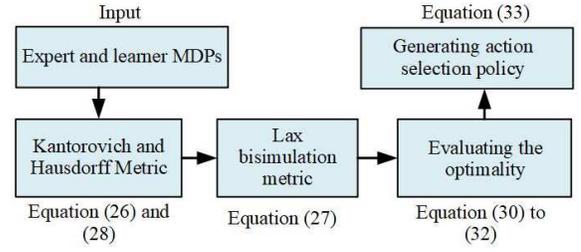}
\caption{Summary of the proposed action selection mapping method.}
\label{fig-rev}
\vspace{-10pt}
\end{figure}

In summary, given $<S_{e},A_{e},T_{e},R_{e}>$ as an expert MDP and $<S_{l},A_{l},$ $T_{l},R_{l}>$ as a learner MDP, $O_{l}(s_{l},a_{l})$ defines the lower bound of potential optimality of $a_{l}$ in terms of the distance between $(s_{l},a_{l})$ and $(s_{e},a^{*}_{e})$. Here $s_{e}$ is considered as the expert state that is closest to $s_{l}$. In this work, we assume experts and learners has the same state definitions, and $s_{e}$ can be easily found accordingly for any $s_{l}$. Finally, the probability of choosing $a_{l}$ is given by:
\begin{equation}\label{eq28}
Pr(a_{l}|s_{l})=\frac{Sig(O_{l}(s_{l},a_{l}))}{\sum\limits_{a\in A_{l}}sig(O_{l}(s_{l},a))},    
\end{equation}
where $Sig$ denotes the Sigmoid function for normalization. Equation (\ref{eq28}) means that actions with higher potential optimality has a higher chance to be selected, and consequently it will improve exploration efficiency.

Finally, we summarize the proposed action selection mapping method in Fig.\ref{fig-rev}. Given the expert and learner MDPs as input, we first calculate the Kantorovich and Hausdorff metrics using equations (\ref{eq22}) and (\ref{eq24}), respectively. Then we calculate the lax bisimulation metric via equation (\ref{eq23}), and evaluate the potential optimality of actions using equations (\ref{eq26}) to (\ref{eq27-1}). 
Consequently, the optimality metrics are normalized, and the action selection probability is produced by applying equation (\ref{eq28}). The proposed QDTRL and ADTRL are summarized in Algorithm 1 and 2, respectively.

\begin{algorithm}[!t]
	\caption{QDTRL-based joint resource allocation}
	\small
	\begin{algorithmic}[1]
		\STATE \textbf{Initialize:} Wireless and QDTRL parameters
		\FOR{$TTI=1$ to $T^{total}$}
		 \FOR{Each BS}
	          	\STATE With probability $\epsilon$, selecting actions randomly; otherwise, choosing actions by $a_{l}=\arg \max\limits_{a}Q(s_{l}',a,w))$. 
	            \STATE GRM implements inter-slice resource allocation as equation (\ref{eq8:main}). 
	        	\STATE SRMs distribute radio resource to UEs by proportional fairness, and replace cached content items. 
		        \STATE Updating system state, and saving $(s_{l},a_{l},r_{l},s_{l}')$ to the experience pool. Every $C$ TTIs, sampling a minbatch from experience pool randomly.
		\STATE Find $\mathcal{F}(s_{l})$ and $\mathcal{F'}(a_{l})$ for any $s_{l}$ and $a_{l}$ in the minibatch.
		\STATE Generating target Q-values $Q^{Tar}(s_{l},a_{l})$= 
		\begin{footnotesize} \begin{equation} \notag 
             \left\{
             \begin{array}{ccl}
            r_{l} &   if\;done\\
            \sigma_{1} Q^{E}(\mathcal{F}(s_{l}),\mathcal{F'}(a_{l}))+ \quad \qquad \qquad\qquad \qquad\qquad  \\ 
             r_{l}+ \gamma Q(s_{l}',\arg \max\limits_{a}Q(s_{l}',a_{l},w),w') &  else  \\
            \end{array} \right.
            \end{equation}\end{footnotesize}
		\STATE Updating $w$ using gradient descent by minimizing the loss $L(w)=Er(Q^{Tar}(s_{l},a_{l})-Q(s_{l},a_{l},w))$.
		\STATE Copying $w$ to $w'$ after several training.
		\ENDFOR
		\ENDFOR
	\STATE \textbf{Output:} Performance of the network and the learning algorithm.	
	\end{algorithmic}
\end{algorithm}

\begin{algorithm}[!t]
	\caption{ADTRL-based joint resource allocation}
	\begin{algorithmic}[1]
		\small
		\STATE \textbf{Initialize:} Wireless and ADTRL parameters
		\FOR{$TTI=1$ to $T^{total}$}
		 \FOR{Each BS}
	          	\STATE With probability $\epsilon$, selecting action $a_{l}$ by using equation (\ref{eq28}); 
Otherwise, choosing $a_{l}$ by $\arg \max\limits_{a}Q(s_{l}',a,w)$.
	            \STATE GRM implements inter-slice resource allocation as equation (\ref{eq8:main}). 
	        	\STATE SRMs distribute radio resource to UEs by proportional fairness, and replace cached content items. 
		        \STATE Updating system state, and saving $(s_{l},a_{l},r_{l},s_{l}')$ to the experience pool. Every $C$ TTIs, sampling a minbatch from experience pool randomly.
		\STATE Generating target Q-values $Q^{Tar}(s_{l},a_{l})$= 
		\begin{small} \begin{equation} \notag 
             \left\{
             \begin{array}{ccl}
            r_{l} &   if\;done\\
             r_{l}+ \gamma Q(s_{l}',\arg \max\limits_{a}Q(s_{l}',a,w),w') &  else  \\
            \end{array} \right.
            \end{equation}\end{small}		
        \STATE Updating $w$ using gradient descent by minimizing the loss $L(w)=Er(Q^{Tar}(s_{l},a_{l})-Q(s_{l},a_{l},w))$.
		\STATE Copying $w$ to $w'$ after several training.
		\ENDFOR
		\ENDFOR
	\STATE \textbf{Output:} Performance of the network and the learning algorithm.	
	\end{algorithmic}
\end{algorithm}

\begin{algorithm}[!t]
	\caption{EB-DQN-based joint resource allocation}
	\small
	\begin{algorithmic}[1]
		\STATE \textbf{Initialize:} Wireless and EB-DQN parameters
		\FOR{$TTI=1$ to $T^{total}$}
		 \FOR{Each BS}
	          	\STATE With probability $\epsilon$, choose actions randomly; otherwise, choosing actions by $a_{l}=\arg \max\limits_{a}Q(s_{l}',a,w))$. 
	            \STATE GRM implements inter-slice resource allocation as equation (\ref{eq8:main}). 
	        	\STATE SRMs distribute radio resource to UEs by proportional fairness, and replace cached content items by TTL. 
		        \STATE Updating system state, and saving $(s_{l},a_{l},r_{l},s_{l}')$ to the experience pool. Every $C$ TTIs, sampling a minbatch from experience pool randomly.
		\STATE Generating target Q-values $Q^{Tar}(s_{l},a_{l})$= 
		\begin{small} \begin{equation} \notag 
             \left\{
             \begin{array}{ccl}
            r_{l}+\frac{\Psi}{\sqrt{\psi(s,a)}} &   if\;done\\
             r_{l}+\frac{\Psi}{\sqrt{\psi(s,a)}}+ \gamma \arg \max\limits_{a}Q(s_{l}',a,w') &  else  \\
            \end{array} \right.
            \end{equation}\end{small}
		\STATE Updating $w$ using gradient descent by minimizing the loss $L(w)=Er(Q^{Tar}(s_{l},a_{l})-Q(s_{l},a_{l},w))$.
		\STATE Copying $w$ to $w'$ after several training.
		\ENDFOR
		\ENDFOR
	\STATE \textbf{Output:} Performance of the network and the learning algorithm.	
	\end{algorithmic}
\end{algorithm}

\begin{algorithm}[!t]
	\small
	\caption{PPF-TTL based joint resource allocation}
	\begin{algorithmic}[1]
		\STATE \textbf{Initialize:} Wireless networks parameters
		\FOR{$TTI=1$ to $T^{total}$}
		\FOR{Each BS}
		\FOR{Each RB}
		 \STATE Calculating the estimated transmission rate of UEs in the queue. 
		 \STATE Calculating proportional fairness metric\cite{b27}.
		 \STATE Transmitting URLLC packets with the highest proportional fairness. If no URLLC packet, then transmitting eMBB packets.
		\ENDFOR
		\STATE BS replaces cached content items by time-to-live rule. 
		\ENDFOR
		\ENDFOR
	\STATE \textbf{Output:} Performance of the network.	
	\end{algorithmic}
\end{algorithm}

\subsection{Baseline: Exploration bonus DQN and PPF-TTL}

In this section, we include two baseline algorithms. Firstly, EB-DQN serves as a benchmark to compare our DTRL method with other ML-based algorithms. The MDP definition of EB-DQN is the same as DTRL, shown in Table \ref{tab1}. EB-DQN agent explores the joint resource allocation task from scratch, and no prior knowledge is included. In EB-DQN, the loss function is defined by:
\begin{equation} \label{eq29}
\resizebox{0.894\hsize}{!}{$\begin{aligned}
L(w)=Er(r+&\frac{\Psi}{\sqrt{\psi(s,a)}}+\gamma \max\limits_{a} Q(s',a,w')-Q(s,a,w)),
\end{aligned}$}
\end{equation}
where $Er$ has been defined in equation (\ref{eq18-11}) as loss function of neural networks, $\Psi$ is an extra reward, and $\psi(s,a)$ is the number of times that $(s,a)$ is selected. $\frac{\Psi}{\sqrt{\psi(s,a)}}$ is regarded as an extra bonus for selecting actions that are less visited, and it encourages the agent to better explore the environment. EB-DQN-based joint resource allocation is summarized in Algorithm 3. 

On the other hand, to compare the ML methods with model-based algorithms, we apply a model-based PPF-TTL algorithm. The well-known priority proportional fairness (PPF) algorithm is applied for radio resource allocation, in which URLLC packets have a higher priority than eMBB packets\cite{b27}. The RBs will first serve URLLC transmission, then eMBB traffic will be processed. We deploy the TTL method for caching, but no slicing and learning are included. The PPF-TTL method is shown in Algorithm 4.

\subsection{Computational Complexity Analyses}

In this section, we analyze the computational complexity of the proposed DTRL-based methods. Firstly, the complexity of the DTRL method is dominated by the training and updating of the LSTM network. The complexity of the LSTM network updating consists of the running time of recurrent connections and bias and the updating time of input and output nodes. The computational complexity for updating the LSTM network in DTRL is $\mathcal{O}(l_{hd}m_{lstm}^{2}c_{lstm}^2)$ \cite{b28-00}, where $l_{hd}$ is the number of hidden layers, $c_{lstm}$ is the number of memory cells in each block and  $m_{lstm}$ is the number of memory blocks. It is worth noting that only the main network needs to be trained, and the target network can copy the weight from the main network.

On the other hand, the knowledge transfer process also contributes to the complexity. In QDTRL, the knowledge transfer consists of the state and action mapping functions $\mathcal{F}$ and $\mathcal{F'}$ (indicated by equation (\ref{eq18-1})). Accordingly, the time complexity is $\mathcal{O}(\sum_{q=1}^{N^e} |S_{e,q}||A_{e,q}|)$, where $|N^{e}|$ is the number of experts, $S_{e,q}$ and $A_{e,q}$ are state and action sets of experts, respectively.

In ADTRL, the Kantorovich distance can be considered as an optimal transportation problem, which is computable within a strong polynomial time $\mathcal{O}(|S|^2log(|S|))$, where $|S|$ is the total number of possible distributions\cite{b28-01}. Meanwhile, the Hausdorff metric can be solved in a nearly linear time\cite{b28-02}, which can be neglected compared with the complexity of Kantorovich distance. Based on equation (\ref{eq27}), the total complexity of knowledge transfer in ADTRL is $\mathcal{O}(|N_{e}||A_{l}||S|^2log(S))$, where $|A_{l}|$ is the action set of the learner. In summary, these analyses show that our knowledge transfer process can be efficiently computed, and the complexity is linearly related to the number of experts.

\section{Performance Evaluation}
\label{s5}
\subsection{Parameter Settings}

In this section, we consider six different cases: expert 1, expert 2, learner 1 (QDTRL), learner 2 (ADTRL), baseline 1 (EB-DQN) and baseline 2 (PPF-TTL). We include 6 adjacent gNBs, and each algorithm is implemented in one gNB randomly. 
Each gNB contains an eMBB slice and a URLLC slice. The eMBB slice has 5 UEs, while the URLLC slice has 10 UEs \cite{b28-03}. The radius of each gNB is 300 meters, and the distance between two adjacent gNBs is 600 meters. For each gNB, there are 100 RBs in total, which are divided into 13 resource block groups (RBGs)\cite{b28}. We assume the caching capacity is reallocated every 50 TTIs because it takes time to replace the cached content items. The experience of experts is presumed to be existing knowledge for learners.

We deploy LSTM networks with 30 nodes as hidden layers for the target and main networks in DTRL and EB-DQN. The network learning rate and the number of layers are selected by the grid search method. We try different parameter combinations and find the best performance accordingly. The simulations include 3000 TTIs, where the first 1500 TTIs are the exploration period, and the remaining TTIs are the exploitation period. The simulations are implemented in MATLAB 5G library with 15 runs to get the average value. Other 5G and learning parameters are shown in Table \ref{tab2}. 

\begin{table}[!tb]
\caption{Parameters settings}
\centering
\renewcommand\arraystretch{1.4}
\begin{tabular}{|p{4cm}<{\centering}||p{3.8cm}<{\centering}|}
\hline
 \textbf{5G settings} & \textbf{Cache settings}\\
\hline
Bandwidth: 20MHz & Caching capacity: 20 items \\ 
3GPP urban macro network & TTL value: 50 TTIs \\ 
Number of RBs: 100 & Contents catalog size: 40/slice \\
  Subcarriers in each RB: 12& Contents popularity: Zipf \\
\cline{2-2}
  Subcarrier bandwidth: 15kHz & \textbf{Traffic model} \\
\cline{2-2}
  Transmission power: 40 dBm (uniform distributed) & URLLC/ eMBB traffic: poisson distribution\\
 TTI size: 2 OFDM symbols &Packet size: 36 Bytes \\ 
  \cline{2-2}
Tx/Rx antenna gain: 15 dB. &\textbf{Learning settings} \\
\hline
\textbf{Retransmission settings}& Network layers: 4\\
\cline{1-1}
 Max number of retransmissions: 1& 2 LSTM network hidden layers \\
 Round trip delay: 4 TTIs & hidden layer has 35 nodes\\
Hybrid automatic repeat request.  & Initial learning rate: 0.005 \\
\cline{1-1}
 \textbf{UE and gNBs} & Experience pool size: 150  \\
\cline{1-1}
25 eMBB UE, 50 URLLC UE & Training frequency: 30 TTIs \\
UE random distribution & Minbatch size: 30\\
Number of gNBs: 5  &  Discount factor: 0.5\\
Inter-gNB distance: 500m &  Epsilon value: 0.05 \\
\cline{1-1}
\textbf{Propagation model} & Reward weight: 0.5\\
\cline{1-1}
128.1+37.6log(distance(km))& Transfer learning rate: 0.7\\
Log-Normal shadowing: 8 dB.& $\Psi$ value for EB-DNQ: 0.5\\
\hline
\end{tabular}
\label{tab2}
\vspace{-10pt}
\end{table}

\begin{figure*}[!t]
\vspace{-5pt}
\centering
\subfigure[Convergence performance comparison of EB-DQN, QDTRL and ADTRL ]{
\includegraphics[width=7cm,height=5.4cm]{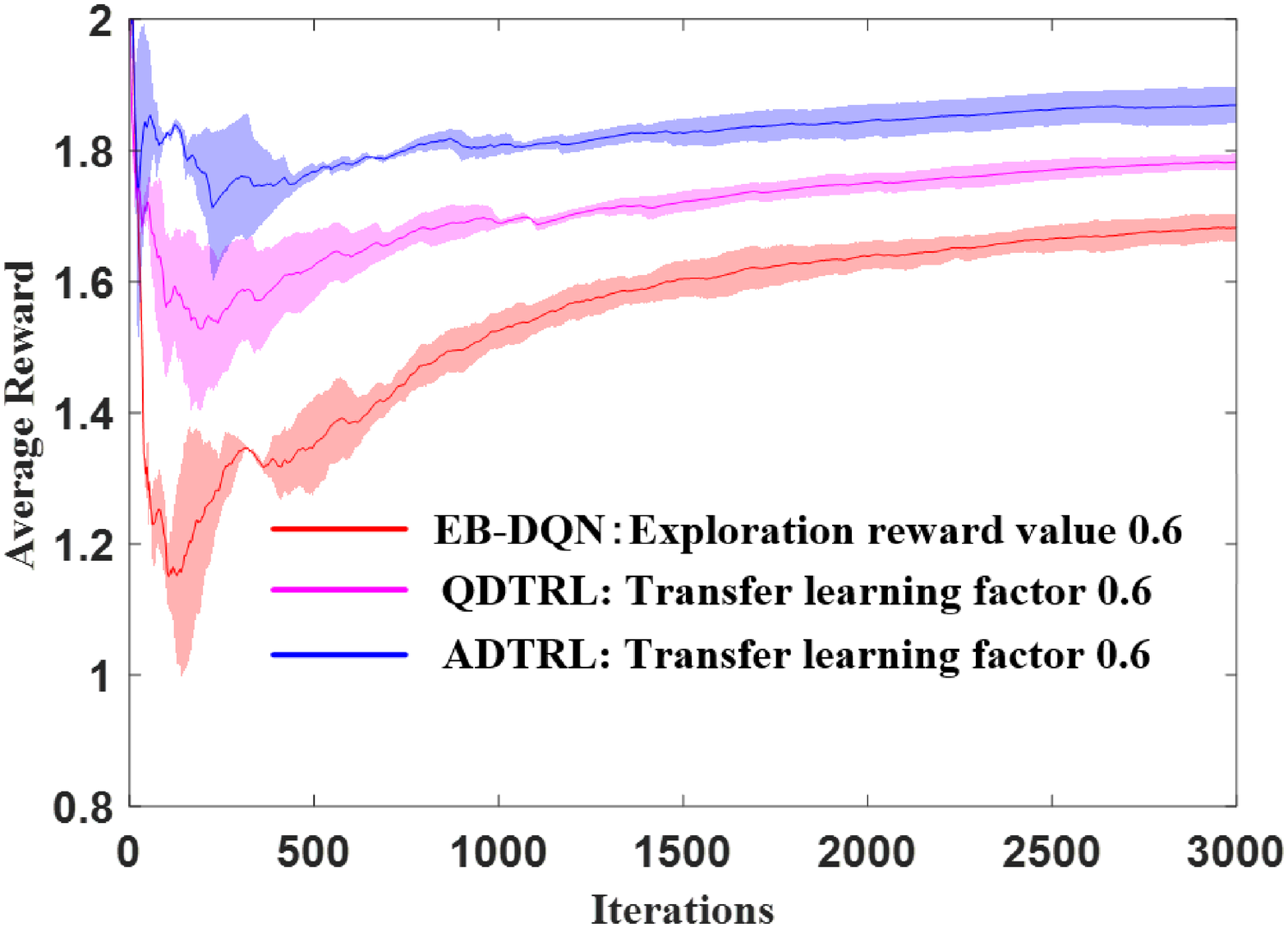}
}
\quad
\subfigure[Network performance of EB-DQN under various extra exploration rewards $\Psi$.   ]{
\includegraphics[width=7cm,height=5cm]{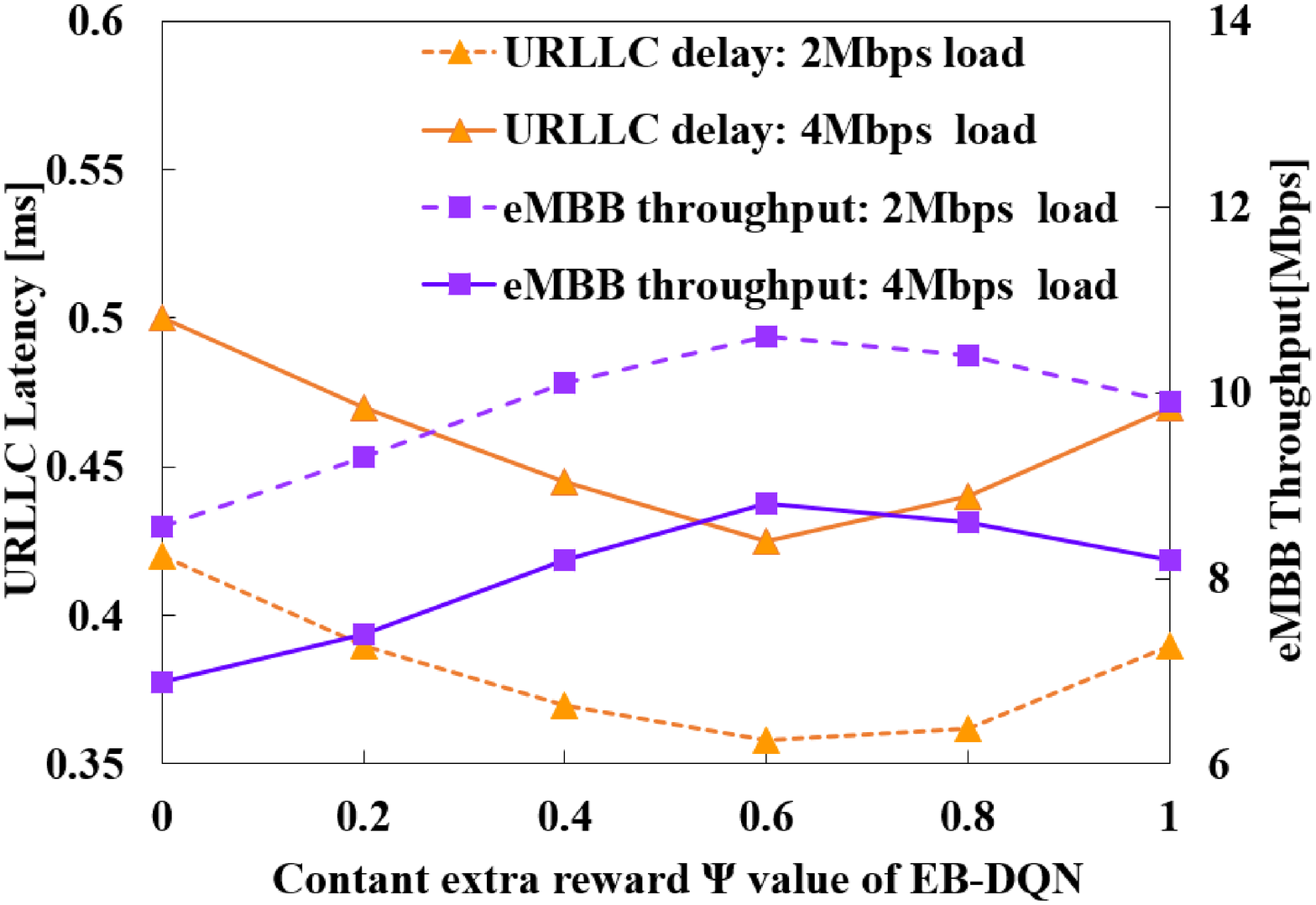}
}
\subfigure[Convergence performance of QDTRL under various Q-table sizes of expert agents (lower value such as 0.3 indicates that learner agent only has 30\% of the expert Q-table)]{
\includegraphics[width=7cm,height=5cm]{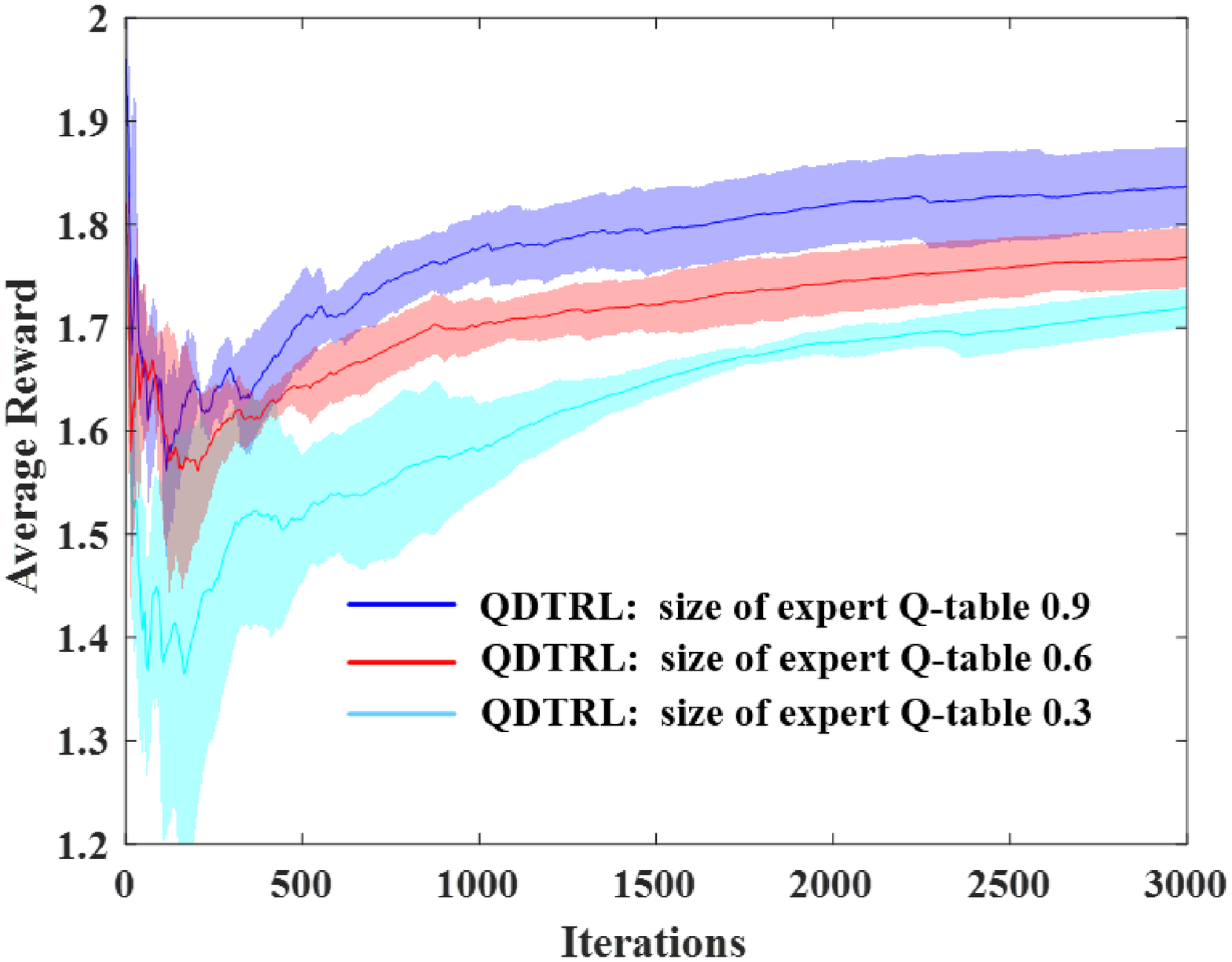}
}
\quad
\subfigure[Convergence performance of ADTRL under various action selection knowledge sizes of expert agents ( the learner agent only has part of the action selection knowledge of expert agents)]{
\includegraphics[width=7cm,height=5cm]{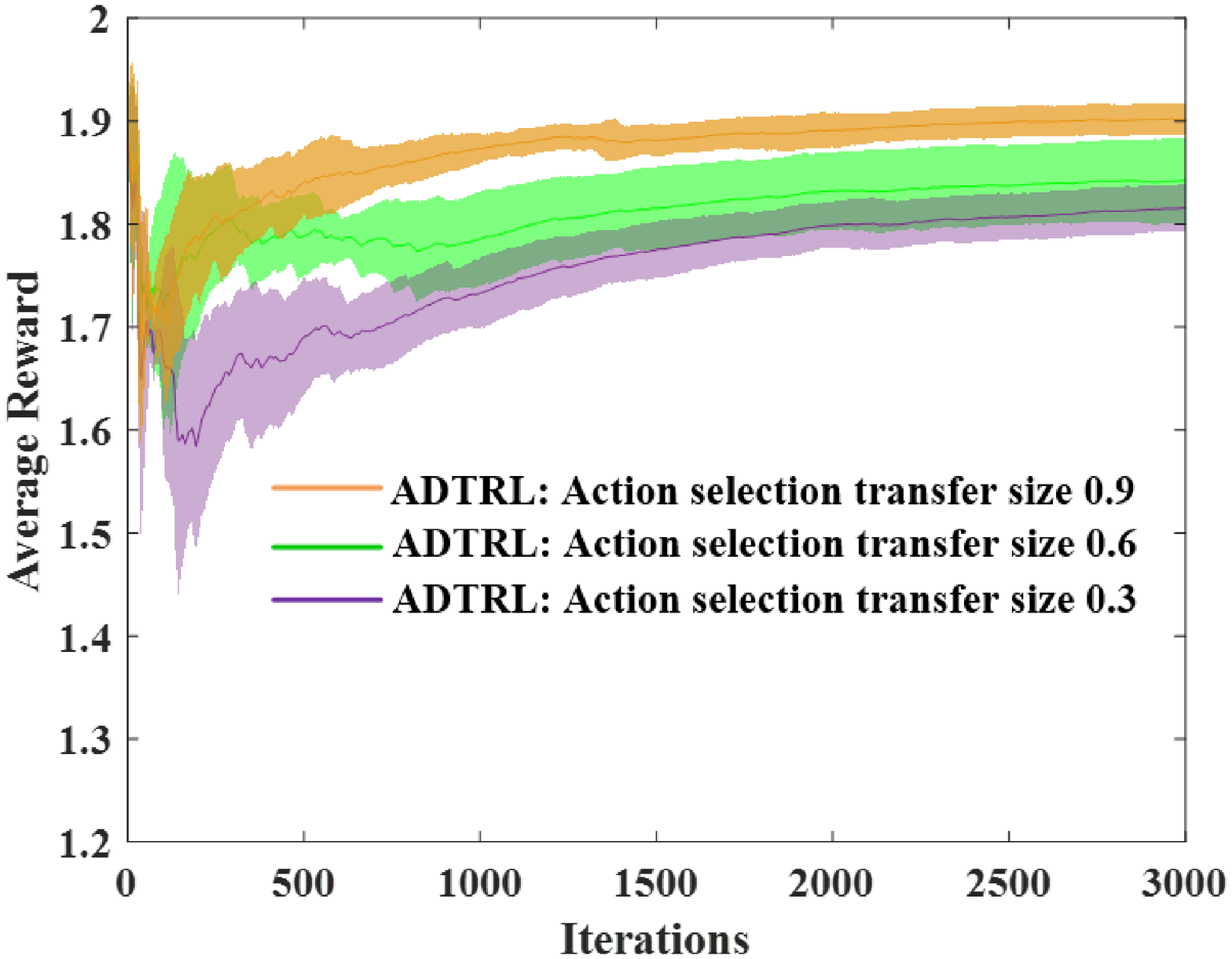}
}
\subfigure[Network performance of QDTRL under various transfer learning rates]{
\includegraphics[width=7cm,height=5cm]{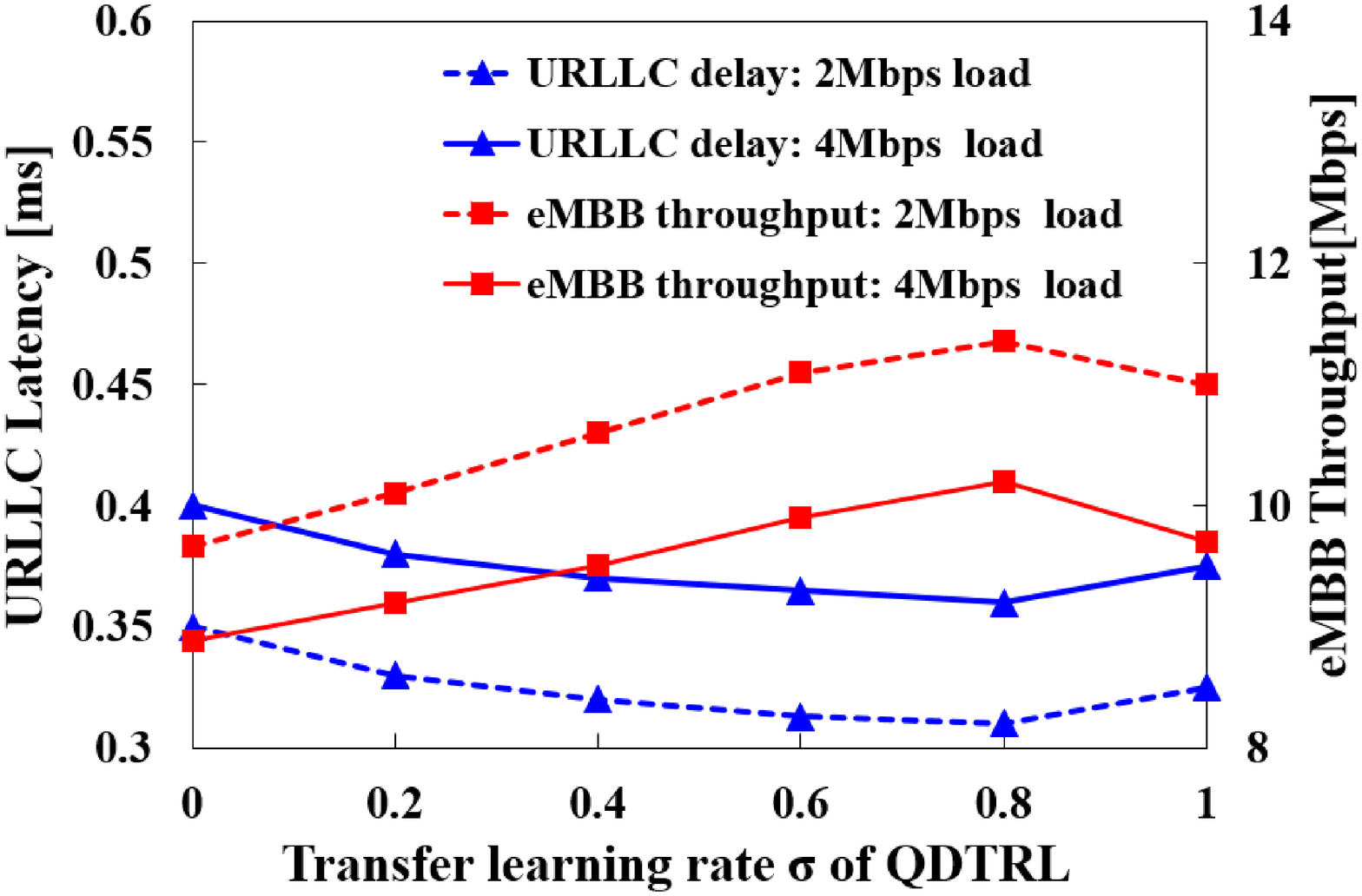}
}
\quad
\subfigure[Network performance of ADTRL under various transfer learning rates]{
\includegraphics[width=7cm,height=5cm]{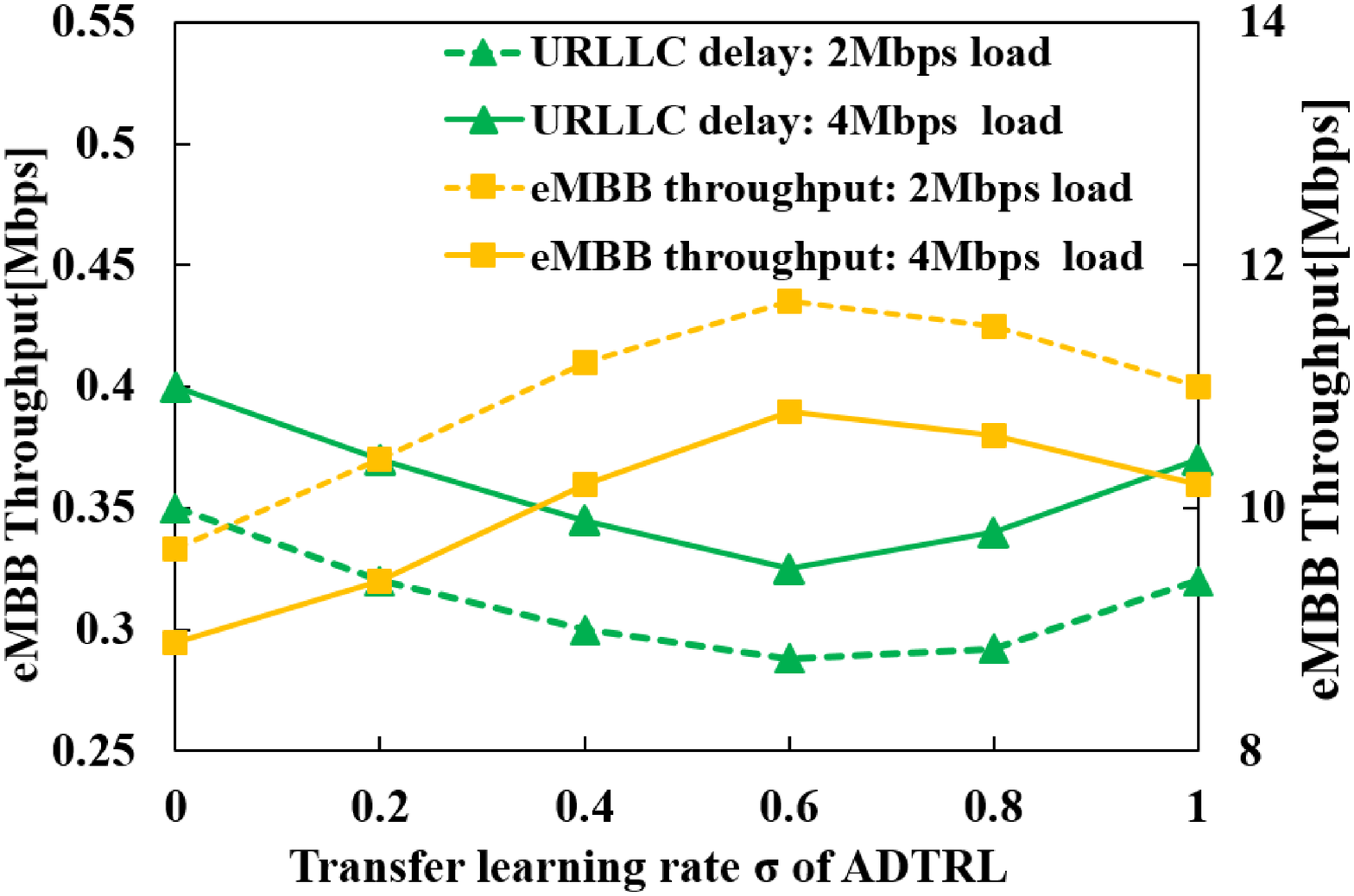}
}
\setlength{\abovecaptionskip}{0pt} 
\caption{Convergence and network performance Comparison against learning parameters.}
\vspace{-5pt}
\label{f5}
\end{figure*}

\subsection{Performance Analyses of Various Learning Parameters}
In this section, we analyze the algorithm performance under diverse learning parameters. Fig.\ref{f5} (a) shows the convergence performance of QDTRL, ADTRL and EB-DQN, which is a critical metric for learning algorithms. ADTRL has the fastest convergence, which can be explained by the improved action selection strategy. ADTRL takes advantage of the action selection policies of experts, which indicates actions with higher potential rewards. It proves that ADTRL applies a more efficient exploration strategy and achieves a better performance. QDTRL also presents a better convergence performance than EB-DQN. In QDTRL, the Q-values of the experts are extracted as extra rewards for action selections. It assumes that actions with higher Q-values in experts can also bring higher rewards to learners, and the exploration is accelerated. However, other actions can still be randomly selected for exploration, lowering the exploration efficiency. On the contrary, in EB-DQN, the agent has no prior knowledge about current tasks. The agent starts from scratch to explore its tasks, which leads to a longer convergence time and a lower average reward.   

Then, Fig.\ref{f5} (b) shows the network performance of EB-DQN against the extra exploration reward $\Psi$ (shown in equation (\ref{eq29})). A higher $\Psi$ value will encourage more explorations, while a lower value means more exploitation. The simulations demonstrate that a higher $\Psi$ may hamper the network performance by over-exploration, and a lower $\Psi$ also degrades the URLLC delay and eMBB throughput by under-exploration. Therefore, an appropriate $\Psi$ value is critical to balance the exploration and exploitation.

\begin{figure*}[!t]
\vspace{-13pt}
\centering
\subfigure[ECDF of URLLC latency (1 Mbps eMBB traffic, 2 Mbps URLLC traffic)]{
\includegraphics[width=7cm,height=5cm]{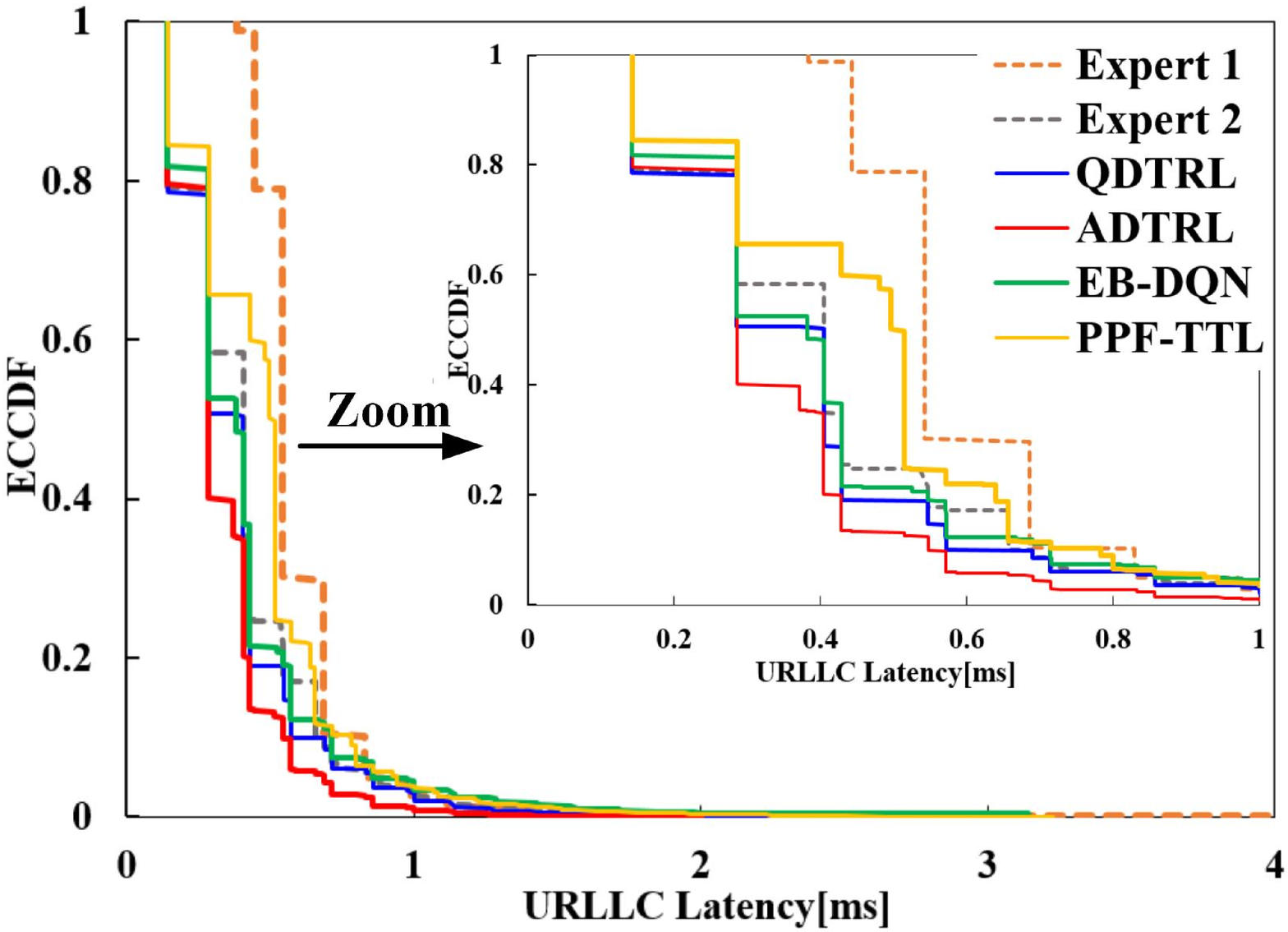}
}
\quad
\subfigure[URLLC latency against traffic load]{
\includegraphics[width=7cm,height=5cm]{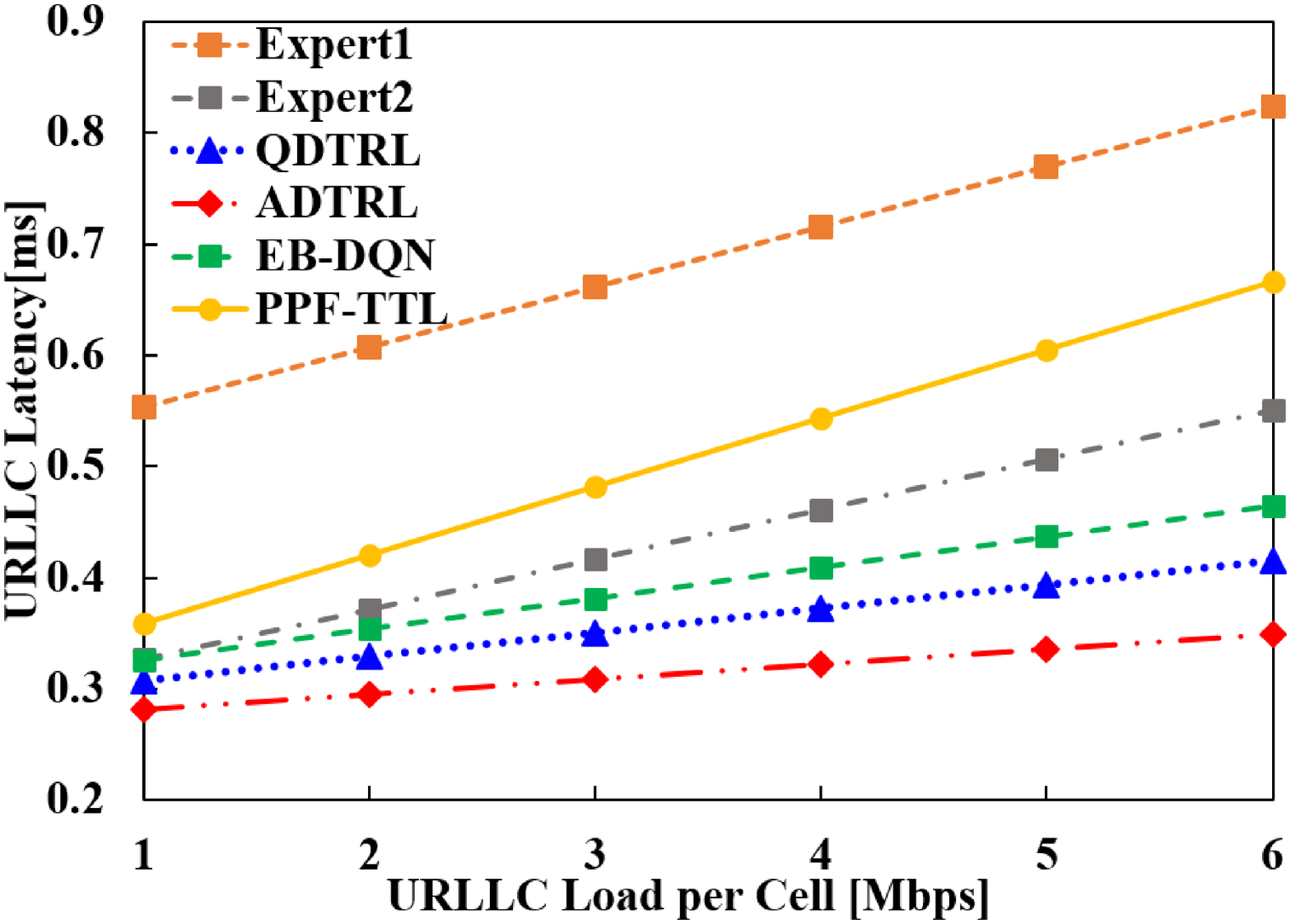}
}
\quad
\subfigure[eMBB throughput against traffic load]{
\includegraphics[width=7cm,height=5cm]{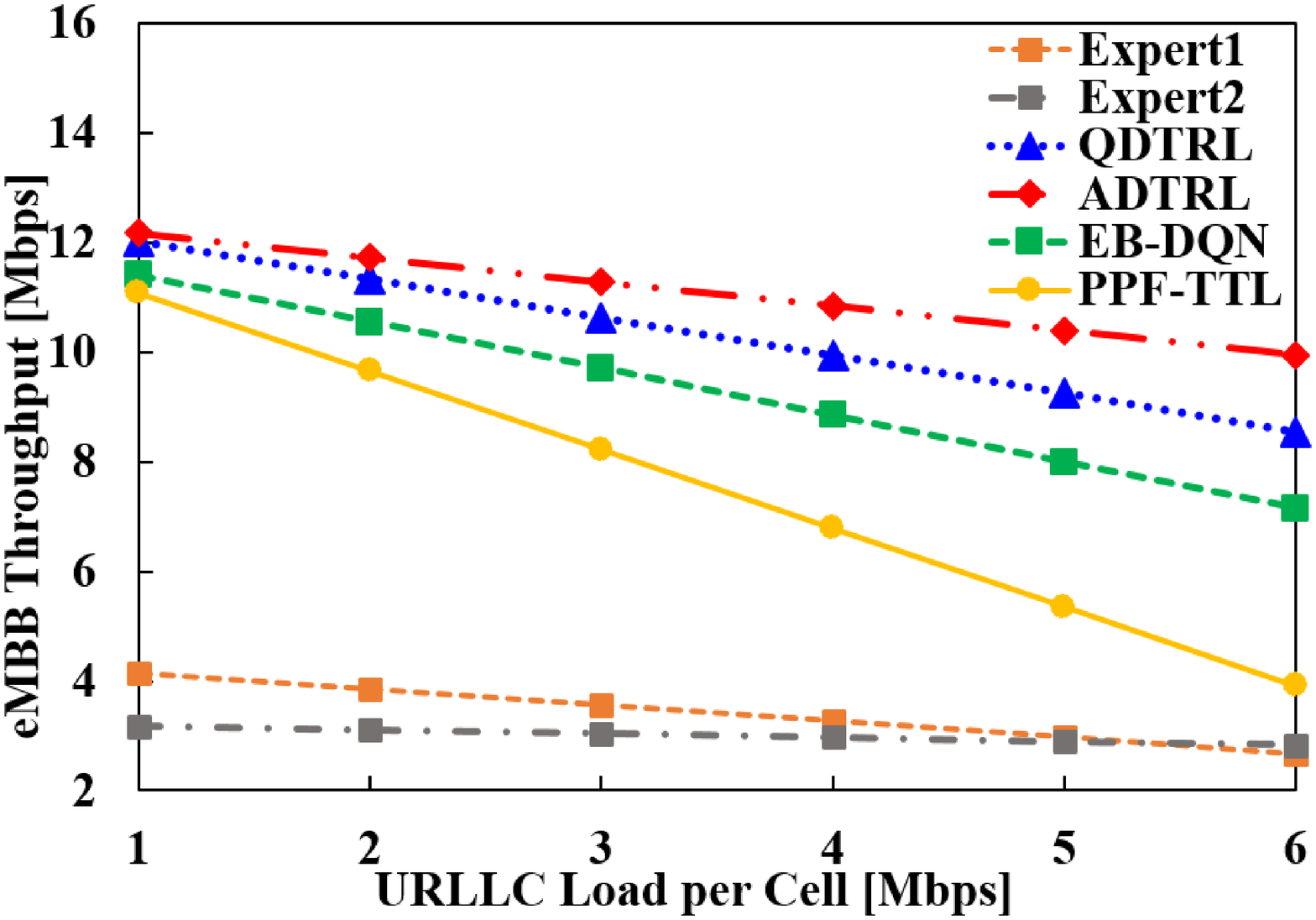}
}
\quad
\subfigure[PDR comparison against traffic load]{
\includegraphics[width=7cm,height=5cm]{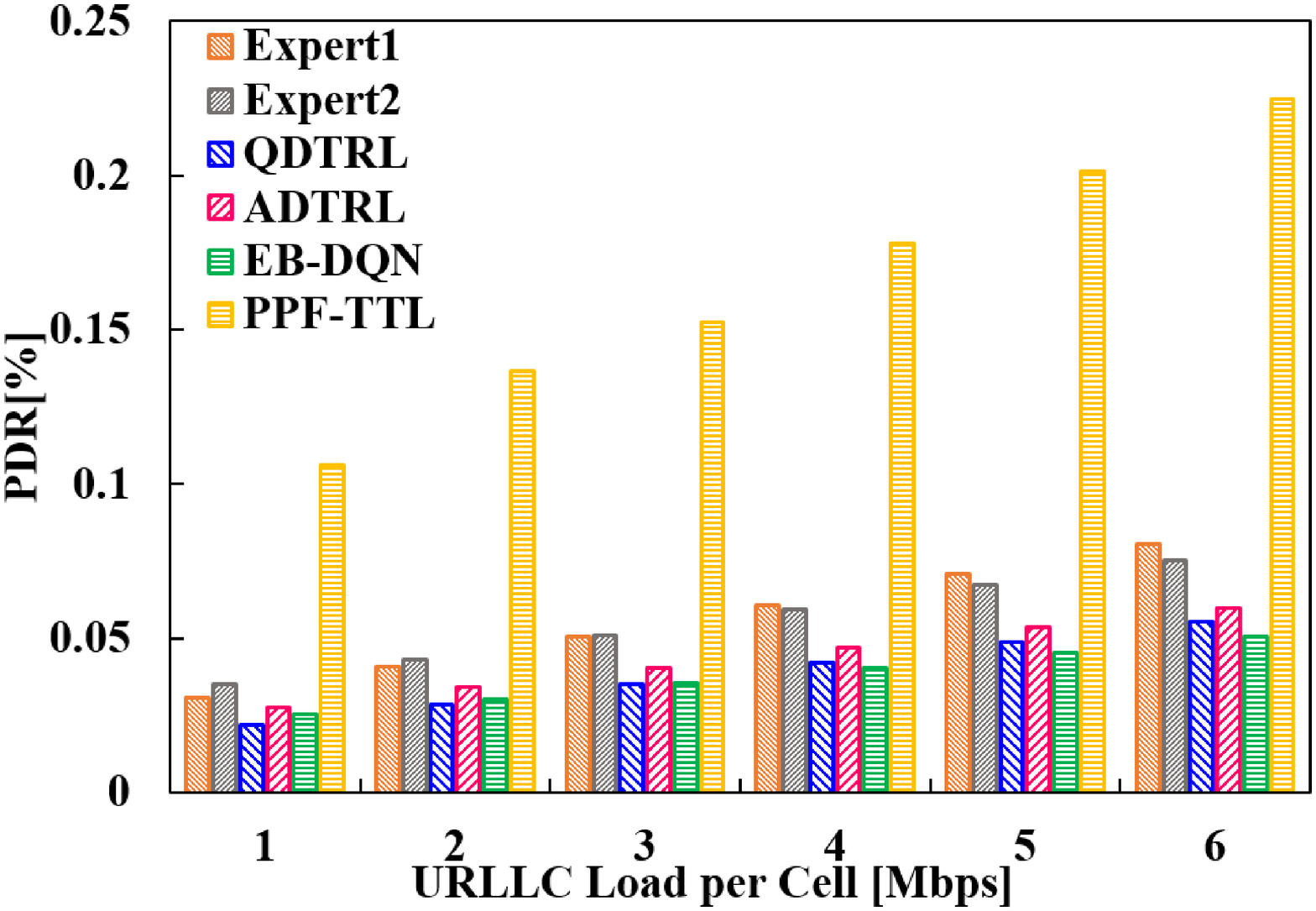}
}
\quad
\subfigure[ URLLC latency against backhaul capacity]{
\includegraphics[width=7cm,height=5cm]{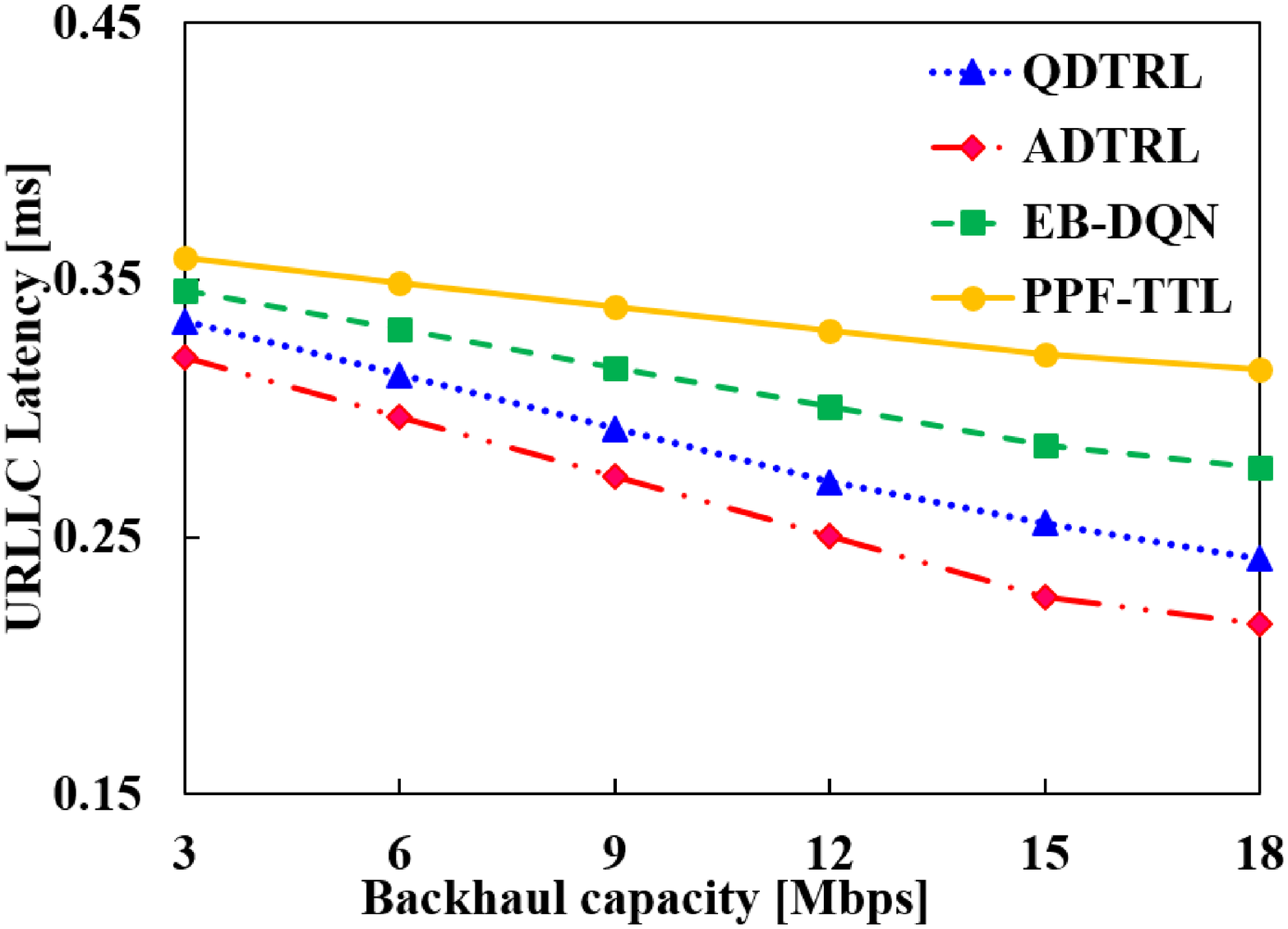}
}
\quad
\subfigure[eMBB throughput against backhaul capacity]{
\includegraphics[width=7cm,height=5cm]{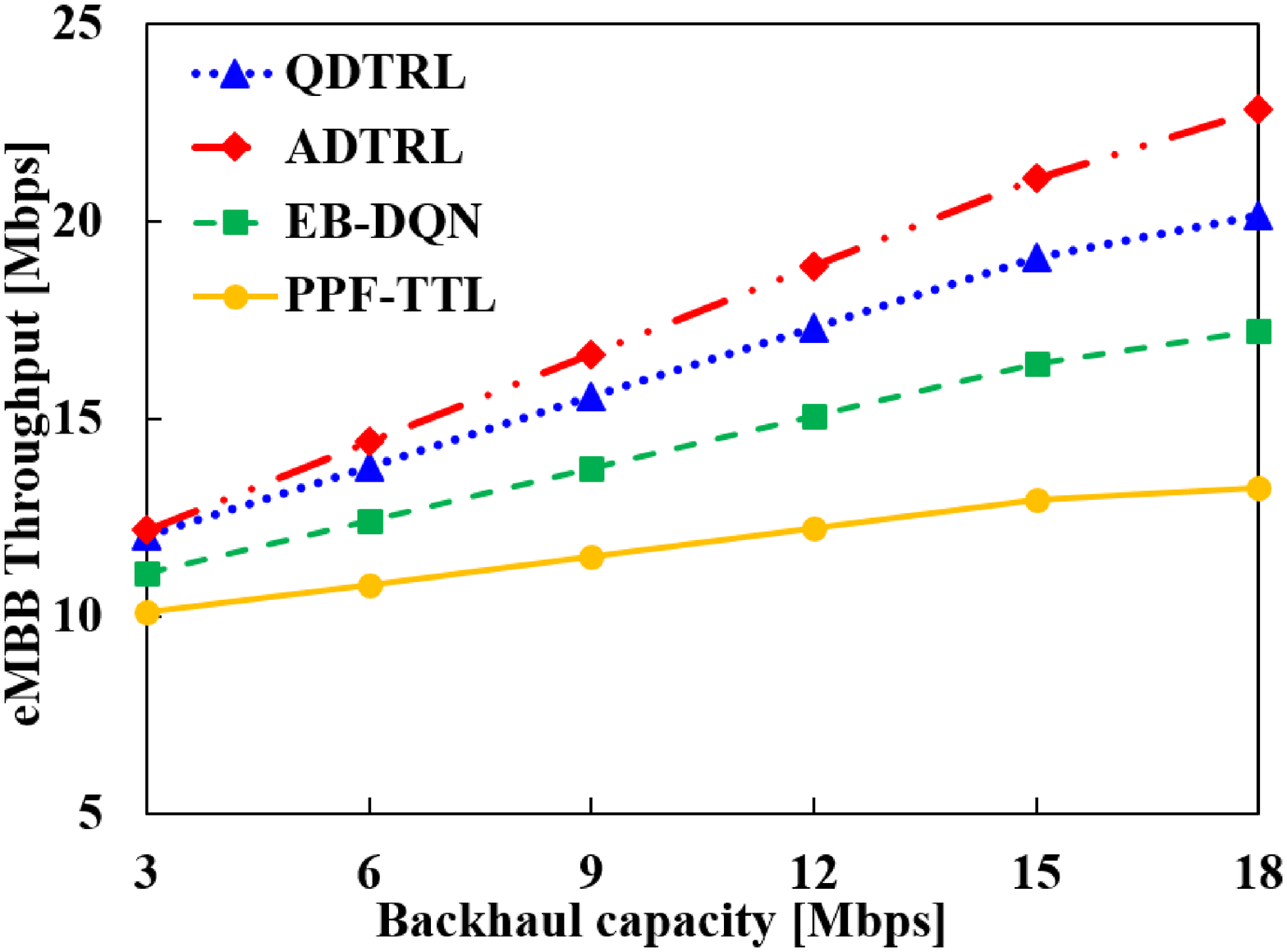}
}
\caption{Performance comparison under various traffic loads and backhaul capacities.}
\label{f2}
\vspace{-13pt}
\end{figure*}

To investigate how the knowledge transfer can contribute to the learner agent performance, Fig. \ref{f5} (c) shows the QDTRL performance under various expert Q-table sizes. In particular, a lower value such as 0.3 means that the learner agent only has 30\% of the expert Q-tables as prior knowledge. Fig. \ref{f5} (c) demonstrates that more prior knowledge can bring better performance for the learner agent, while partial prior knowledge may lower average rewards. Similarly, Fig. \ref{f5} (d) presents the ADTRL performance using different action selection transfer sizes. Specifically, a lower value indicates that the learner agent only has part of the action selection knowledge of expert agents. Consequently, one can observe that more transferred knowledge can improve exploration efficiency and produce a higher average reward for the learner agent.

Finally, note that we have defined transfer learning rates when introducing QDTRL and ADTRL, which represents the importance of transferred knowledge. The network performance of QDTRL and ADTRL under different transfer learning rates is investigated here. In QDTRL, the transfer learning rate is indicated by the $\sigma_{1}$ in equation (\ref{eq20}). Fig.\ref{f5} (e) shows that a higher transfer learning rate may lead to better network performance, which is indicated by a lower URLLC delay and a higher eMBB throughput. However, a very high transfer learning rate may affect the exploration of the agent itself, and it leads to sub-optimal results such as higher delays and lower throughput. A similar trend can be observed in Fig.\ref{f5} (f) for ADTRL, in which the transfer learning rate is indicated by variable $\sigma_{2}$ in equation (\ref{eq27-1}). A higher $\sigma_{2}$ value significantly reduces the exploration complexity and brings better network performance. However, when $\sigma_{2}\geq0.6$, the transferred knowledge dominates the learning process, and it results in performance degradation in terms of latency and throughput.

\subsection{Network Performance Analyses}
\label{s5-1}

In this section, we compare the network performance of different algorithms under various traffic loads and backhaul capacities. The eMBB traffic is fixed to 1 Mbps per cell, and the URLLC traffic ranges from 1 to 6 Mbps. We first present the results, then explain the performance of each algorithm.

Fig.\ref{f2} (a) shows the Empirical Complementary Cumulative Distribution Function (ECCDF) of the URLLC latency with 2 Mbps URLLC traffic, which presents empirical distributions of packet delays. We zoom the area where the ECCDF value is higher than 0.1 and the URLLC delay is lower than 1 ms to better show the results. Expert 1 and the PPF-TTL method present the highest delay, which is indicated by the high delay distribution in 0.1-1 interval of the ECCDF axis. By contrast, expert 2 has a lower delay distribution than expert 1. Meanwhile, EB-DQN and ADTRL show comparable delay performance for the URLLC slice. Finally, ADTRL achieves the best delay performance than other algorithms. 
Fig.\ref{f2} (b) and (c) present average URLLC slice delay and eMBB slice throughput against traffic loads. It shows that both ADTRL and QDTRL maintain lower delay and higher throughput than experts and baseline algorithms under various traffic loads. Expert 2 shows a lower URLLC delay than the PPF-TTL and expert 1, but its eMBB throughput is much lower than any other algorithm. In Fig.\ref{f2} (d), all algorithms have comparable packet drop rate (PDR) except the PPF-TTL. In the following, we will explain the results of each algorithm.

We first analyze the experts' performance. The Expert 1 shows a high delay and a low throughput, because it has no caching capability. All packets required by UEs have to be processed by the core network, and the constant backhaul delay leads to a high URLLC delay and a low eMBB throughput. On the contrary, expert 2 has a lower delay because we apply a fixed RB allocation strategy. In particular, RBs are distributed according to the UE numbers in each slice; thus the URLLC slice always has more RBs, which leads to a low URLLC delay. But the eMBB slice is affected by a low eMBB throughput. 

For the baseline algorithms, EB-DQN outperforms the PPF-TTL method because it jointly learns the radio and cache resource allocation. The eMBB throughput of the PPF-TTL method decreases significantly with the increasing URLLC load, which results from the priority settings in this method. In the PPF method, whenever a new URLLC packet arrives, it is directly scheduled over the eMBB packets, which unavoidably degrades the eMBB throughput.  

Finally, the proposed QDTRL and ADTRL have the best overall performance. Due to the novel DTRL-based scheme, they can leverage the knowledge of experts and further improve their own performance on new tasks. When the URLLC traffic is 4 Mbps, ADTRL presents a 21.4\% lower URLLC delay and a 22.4\% higher eMBB throughput than EB-DQN. A 40.8\% lower URLLC delay and a 59.8\% higher eMBB throughput are also observed compared with the PPF-TTL method. The simulations show that the proposed DTRL-based solutions achieve more promising results than the baseline algorithms.  

Moreover, backhaul capacity is one of the main bottlenecks of 5G RAN. Here we investigate the network performance under various backhaul capacities, and the results are shown in Fig.\ref{f2} (e) and (f). Note that the basic performance of experts has been shown in former results, and here we mainly compare the DTRL-based solutions with baseline algorithms. 

As expected, all algorithms achieve lower delays for URLLC slice and higher throughput for eMBB slice with increasing backhaul capacity, because higher capacity will reduce the backhaul delay. Moreover, QDTRL and ADTRL still achieve lower URLLC delays and higher eMBB throughput. Compared with EB-DQN, the satisfying performance of QDTRL and ADTRL can still be explained by their knowledge transfer strategy. Meanwhile, the worst performance of PPF-TTL shows that learning-based methods outperform model-based algorithms by the superior learning capability. When the backhaul capacity is 15 Mbps, ADTRL presents an 18.9\% lower URLLC delay and a 24.2\% higher eMBB throughput than EB-DQN. Compared with the PPF-TTL method, a 24.7\% lower URLLC delay and a 54.3\% higher eMBB throughput are observed.

\begin{figure}[!t]
\centering
\subfigure[ URLLC latency against caching capacity]{
\includegraphics[width=7cm,height=5cm]{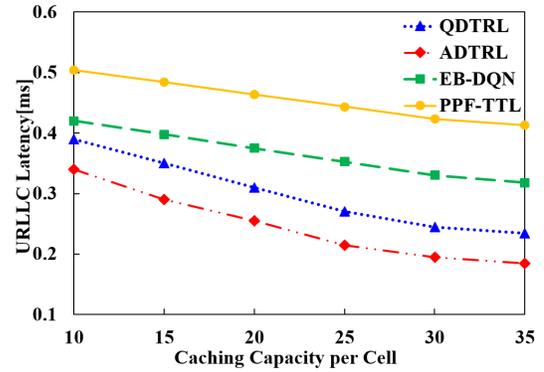}
}
\subfigure[eMBB throughput against caching capacity]{
\includegraphics[width=7cm,height=5cm]{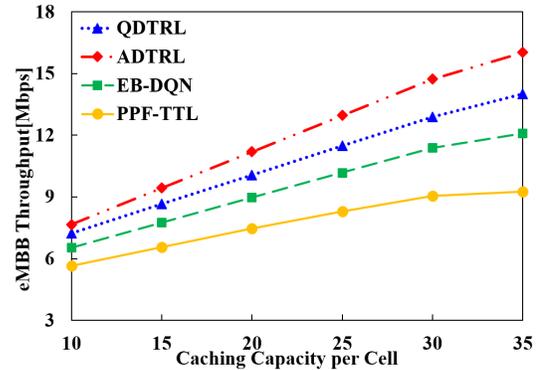}
}
\subfigure[Cached hit ratio against caching capacity ]{
\includegraphics[width=7cm,height=5cm]{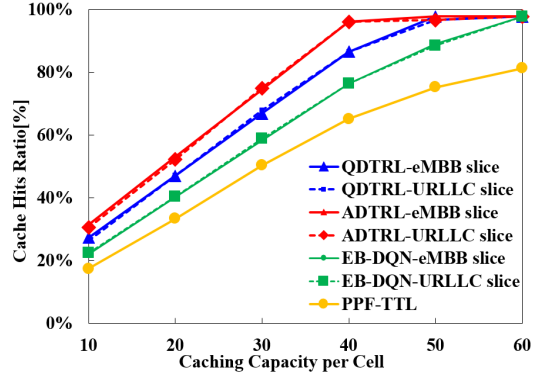}
}
\setlength{\abovecaptionskip}{0pt} 
\caption{Network performance comparison under various caching capacities.}
\vspace{-10pt}
\label{f4}
\end{figure}

\subsection{Content Caching Performance Analyses }
In this section, we compare different algorithms under various caching capacities, which is indicated by the maximum number of content items that can be stored in gNBs. As shown in Fig.\ref{f4} (a) and (b), a higher caching capacity will reduce the URLLC latency and increase the eMBB throughput. It is because a higher caching capacity means that more items can be cached in gNBs, and the average backhaul delay will be reduced. The simulations show that ADTRL and QDTRL have the best overall performance. EB-DQN still achieves a lower URLLC delay and a higher eMBB throughput than the PPF-TTL algorithm.

Furthermore, we present the cache hit ratio of eMBB and URLLC slices in Fig.\ref{f4} (c). The cache hit ratio represents the proportion of packets that can be found in the cache server when they are required. A higher cache hit ratio usually indicates better network performance, which is affected by the caching capacity and the content replacement strategy. With the increasing caching capacity, cache hit ratios naturally increase for all algorithms. As expected, the cache hit ratio of the eMBB and URLLC slices are well maintained in QDTRL and ADTRL. Compared with ADTRL, the ratios in EB-DQN and PPF-TTL methods are 19.8\% and 31\% lower. Note that the PPF-TTL method only has one curve because we assume there is no slicing in the PPF-TTL algorithm. Finally, ADTRL, QDTRL and EB-DQN have similar cache hit ratios when the caching capacity is 60. It means that most required items can be cached with this capacity, and then a high cache hit ratio is observed.

\section{Conclusion}
\label{s6}
Network slicing is a key technique to enhance flexibility in 5G networks, and ML techniques offer promising solutions. Although widely used reinforcement learning techniques have yielded to improved network performance, they suffer from a long convergence time and lack of generalization. In that sense, knowledge transfer emerges as an important approach to improve learning performance. Yet, transfer learning in wireless has been explored only very recently and in very few studies. This work presented two novel deep transfer reinforcement learning-based solutions for the joint radio and cache resource allocation. The proposed algorithms have been compared with two baseline algorithms via simulations. These results have shown that the proposed methods achieve better network performance and faster convergence speeds than these other benchmarks. In the future, we plan to consider the knowledge transfer between tasks with different state definitions.     

\appendix
In the following we prove equation (10). Recalling equation (8) and (9)
\begin{equation}\nonumber
h_{j,n,g}=\sum_{m\in \mathcal{M}_{j,n}}\phi_{j,n,m,g}T_{j,n,m,g},
\end{equation}
\begin{equation}\nonumber
\sum _{g\in \mathcal{G}_{n}}\sum _{m\in \mathcal{M}_{n}}h_{j,n,m,g}=\frac{C_{j,n}}{C_{j,T}}.
\end{equation}
Then we have 
\begin{equation}\nonumber
h_{j,n,g}\sum _{g\in \mathcal{G}_{n}}\sum _{m\in \mathcal{M}_{n}}h_{j,n,m,g}=\frac{C_{j,n}}{C_{j,T}}\sum_{m\in \mathcal{M}_{j,n}}\phi_{j,n,m,g}T_{j,n,m,g},
\end{equation}
which can be easily transformed to
\begin{equation}\nonumber
\begin{aligned}
h_{j,n,g}&=\frac{C_{j,n}}{C_{j,T}}\frac{\sum_{m\in \mathcal{M}_{j,n}}\phi_{j,n,m,g}T_{j,n,m,g}}{\sum _{g\in \mathcal{G}_{n}}\sum _{m\in \mathcal{M}_{n}}h_{j,n,m,g}}\\
&=\frac{C_{j,n}}{C_{j,T}}\frac{T_{j,n,g}\sum_{m\in \mathcal{M}_{j,n}}\phi_{j,n,m,g}}{\sum _{g\in \mathcal{G}_{n}}\sum _{m\in \mathcal{M}_{n}}h_{j,n,m,g}} \\
&\text{\qquad \qquad \quad(using $T_{j,n,m,g}=T_{j,n,g}$ assumption)}\\
&=\frac{C_{j,n}}{C_{j,T}} \frac{T_{j,n,g}\sum_{m\in \mathcal{M}_{j,n}}\phi_{j,n,m,g}}{\sum _{g\in \mathcal{G}_{n}}\sum _{m\in \mathcal{M}_{n}}\phi_{j,n,m,g}T_{j,n,m,g}}\\
&\text{\qquad \qquad \quad(using $h_{j,n,m,g}=\phi_{j,n,m,g}T_{j,n,m,g}$)}\\
&=\frac{C_{j,n}}{C_{j,T}} \frac{\sum_{m\in \mathcal{M}_{j,n}}\phi_{j,n,m,g}}{\sum _{g\in \mathcal{G}_{n}}\sum _{m\in \mathcal{M}_{n}}\phi_{j,n,m,g}}\\
&=\frac{C_{j,n}}{C_{j,T}} \frac{\sum_{m\in \mathcal{M}_{j,n}}\phi_{j,n,m,g}}{\sum _{m\in \mathcal{M}_{n}}\phi_{j,n,m}},
\end{aligned}
\end{equation}
which is the equation (10).

\section*{Acknowledgment}
We would like to thank Dr. Medhat Elsayed for initial discussions on transfer learning.

\vspace{-40pt}
\begin{IEEEbiography}[{\includegraphics[width=1in,height=1.2in,clip,keepaspectratio]{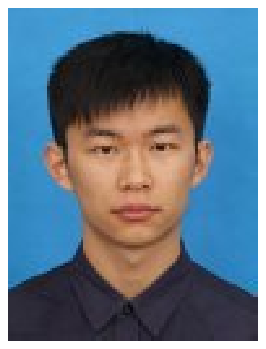}}]{Hao Zhou} is a Phd candidate at the University of Ottawa. He got his B.Eng. and M.Eng degrees from Huazhong University of Science and Technology in 2016, and Tianjin University in 2019, respectively, in China. He is working towards his Phd degree at the University of Ottawa since Sep. 2019. His research interests include electric vehicles, microgrid energy trading,  resource management and network slicing in 5G. He is devoted to applying machine learning techniques for smart grid and 5G applications.     
\end{IEEEbiography}
\vspace{-40pt}
\begin{IEEEbiography}[{\includegraphics[width=1in,height=1.2in,clip,keepaspectratio]{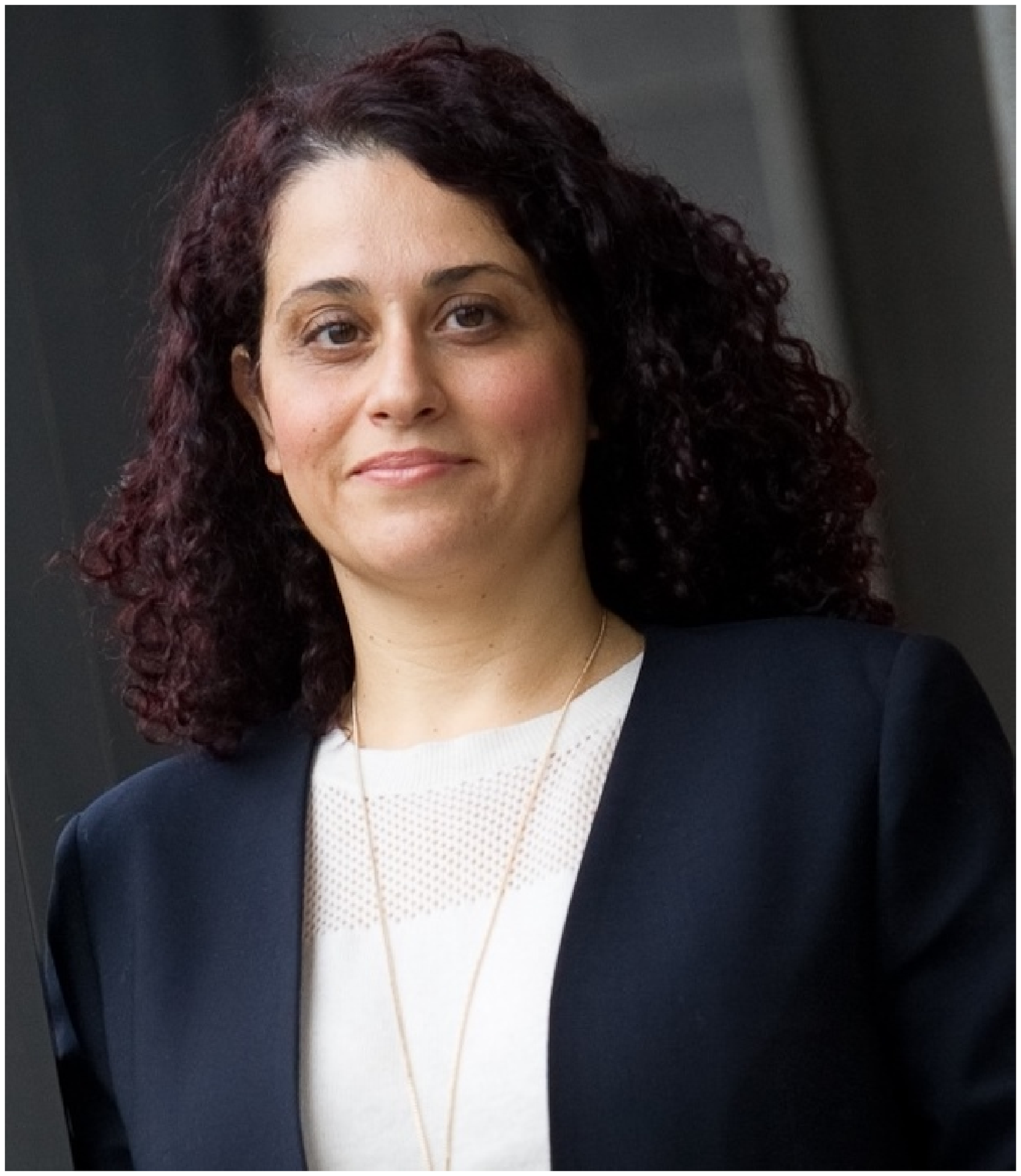}}] {Melike Erol-Kantarci} is Canada Research Chair in AI-enabled Next-Generation Wireless Networks and Associate Professor at the School of Electrical Engineering and Computer Science at the University of Ottawa. She is the founding director of the Networked Systems and Communications Research (NETCORE) laboratory. She has received numerous awards and recognitions. Dr. Erol-Kantarci is the co-editor of three books on smart grids, smart cities and intelligent transportation. She has over 180 peer-reviewed publications. She has delivered 70+ keynotes, plenary talks and tutorials around the globe. She is on the editorial board of the IEEE Transactions on Cognitive Communications and Networking, IEEE Internet of Things Journal, IEEE Communications Letters, IEEE Networking Letters, IEEE Vehicular Technology Magazine and IEEE Access. She has acted as the general chair and technical program chair for many international conferences and workshops. Her main research interests are AI-enabled wireless networks, 5G and 6G wireless communications, smart grid and Internet of Things. She is an IEEE ComSoc Distinguished Lecturer, IEEE Senior member and ACM Senior Member.
\end{IEEEbiography}
\vspace{-40pt}
\begin{IEEEbiography}[{\includegraphics[width=1in,height=1.2in,clip,keepaspectratio]{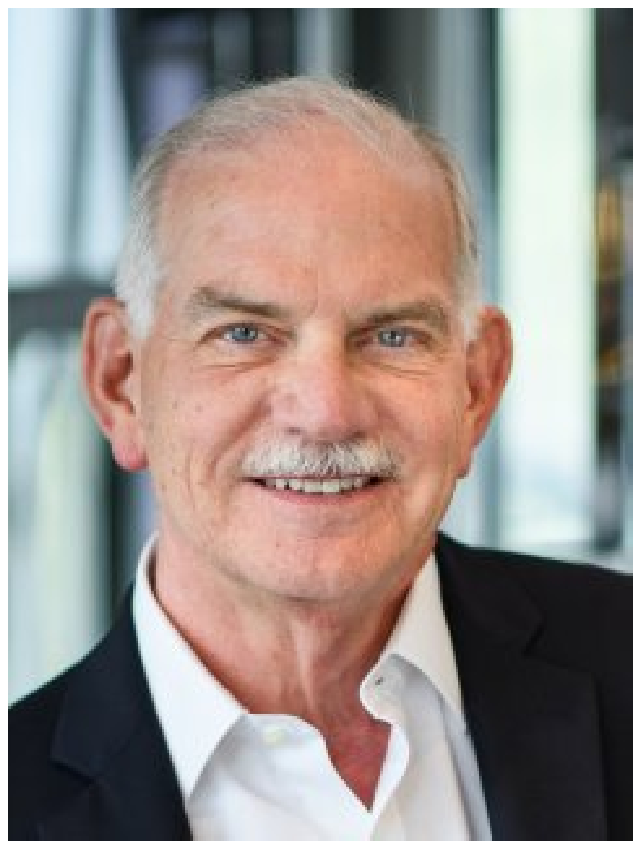}}]{H. Vincent Poor} (S’72, M’77, SM’82, F’87) received the Ph.D. degree in EECS from Princeton University in 1977.  From 1977 until 1990, he was on the faculty of the University of Illinois at Urbana-Champaign. Since 1990 he has been on the faculty at Princeton, where he is currently the Michael Henry Strater University Professor. During 2006 to 2016, he served as the dean of Princeton’s School of Engineering and Applied Science. He has also held visiting appointments at several other universities, including most recently at Berkeley and Cambridge. His research interests are in the areas of information theory, machine learning and network science, and their applications in wireless networks, energy systems and related fields. Among his publications in these areas is the forthcoming book Machine Learning and Wireless Communications.  (Cambridge University Press). Dr. Poor is a member of the National Academy of Engineering and the National Academy of Sciences and is a foreign member of the Chinese Academy of Sciences, the Royal Society, and other national and international academies. He received the IEEE Alexander Graham Bell Medal in 2017.     
\end{IEEEbiography}

\end{document}